\documentclass[10pt,letterpaper]{article}
\usepackage[top=0.85in,left=2.75in,footskip=0.75in]{geometry}

\usepackage{changepage}
\usepackage[utf8]{inputenc}
\usepackage{textcomp,marvosym}
\usepackage{fixltx2e}
\usepackage{amsmath,amssymb}
\usepackage{cite}
\usepackage{nameref}
\usepackage{microtype}
\DisableLigatures[f]{encoding = *, family = * }
\usepackage{rotating}
\usepackage[aboveskip=1pt,labelfont=bf,labelsep=period,justification=raggedright,
  singlelinecheck=off]{caption}

\usepackage{setspace} 
\raggedright
\makeatletter
\renewcommand{\@biblabel}[1]{\quad#1.}
\makeatother

\date{}

\usepackage{lastpage,fancyhdr,graphicx}
\usepackage{epstopdf}
\fancyheadoffset[L]{0.25in}
\fancyfootoffset[L]{0.25in}
\usepackage{ragged2e}
\justifying
\usepackage[numbers]{natbib}
\usepackage{amsmath}
\usepackage{amssymb}
\usepackage{graphicx}
\usepackage{multirow}

\newcommand{\VAR}{\operatorname{VAR}}
\newcommand{\COR}{\operatorname{COR}}

\newcommand{\F}{F_{\mathrm{ST}}}
\newcommand{\hF}{\hat{F}_{\mathrm{ST}}}
\newcommand{\rP}{\hat\rho_{\mathrm{P}}}
\newcommand{\rD}{\hat\rho_{\mathrm{D}}}
\newcommand{\rL}{\hat\rho_{\mathrm{L}}}
\newcommand{\rCV}{\hat\rho_{\mathrm{CV}}}
\newcommand{\rb}{\bar\rho}
\newcommand{\nTR}{n_{\mathrm{TR}}}
\newcommand{\nTA}{n_{\mathrm{TA}}}
\newcommand{\km}{\textit{k}-means}
\newcommand{\bk}{\bar{k}}
\newcommand{\pkg}[1]{\textbf{#1}}
\newcommand{\ls}{\leqslant}
\newcommand{\gs}{\geqslant}

\begin{document}
\pagestyle{fancy}
\vspace*{0.35in}

\begin{flushleft}
{\Large
\textbf\newline{Using Genetic Distance to Infer the Accuracy of Genomic
  Prediction - Submission to PLOS Journals}
}
\newline
% Insert author names, affiliations and corresponding author email (do not include titles, positions, or d
\\
Marco Scutari\textsuperscript{1},
Ian Mackay\textsuperscript{2},
David Balding\textsuperscript{3,4*,\textcurrency}
\\
\bigskip
\bf{1} Department of Statistics, University of Oxford, Oxford, United Kingdom
\\
\bf{2} National Institute of Agricultural Botany (NIAB), Cambridge, United
       Kingdom
\\
\bf{3} Centre for Systems Genomics, School of BioSciences and of Mathematics \&
       Statistics, University of Melbourne, Melbourne, Australia
\\
\bf{4} Genetics Institute, University College London (UCL), London, United Kingdom
\bigskip

% Current address notes
\textcurrency{} Centre for Systems Genomics, Royal Parade, University of
    Melbourne Vic 3010 Australia 

% Use the asterisk to denote corresponding authorship and provide email address in note below.
* d.balding@ucl.ac.uk

\end{flushleft}

\section*{Abstract}

The prediction of phenotypic traits using high-density genomic data has many 
applications such as the selection of plants and animals of commercial interest;
and it is expected to play an increasing role in medical diagnostics. 
Statistical models used for this task are usually tested using cross-validation,
which implicitly assumes that new individuals (whose phenotypes we would like
to predict) originate from the same population the genomic prediction model is
trained on.

In this paper we propose an approach based on clustering and resampling to
investigate the effect of increasing genetic distance between training and
target populations when predicting quantitative traits. This is important for
plant and animal genetics, where genomic selection programs rely on the
precision of predictions in future rounds of breeding. Therefore, estimating
how quickly predictive accuracy decays is important in deciding which training
population to use and how often the model has to be recalibrated. We find that
the correlation between true and predicted values decays approximately linearly
with respect to either $\F$ or mean kinship between the training and the target
populations. We illustrate this relationship using simulations and a collection
of data sets from mice, wheat and human genetics.

\section*{Author Summary}

The availability of increasing amounts of genomic data is making the use of
statistical models to predict traits of interest a mainstay of many applications
in life sciences. Applications range from medical diagnostics for common and 
rare diseases to breeding characteristics such as disease resistance in plants
and animals of commercial interest. We explored an implicit assumption of how
such prediction models are often assessed: that the individuals whose traits we
would like to predict originate from the same population as those that are used
to train the models. This is commonly not the case, especially in the case of
plants and animals that are parts of selection programs. To study this problem
we proposed a model-agnostic approach to infer the accuracy of prediction models
as a function of two common measures of genetic distance. Using data from plant,
animal and human genetics, we find that accuracy decays approximately linearly
in either of those measures. Quantifying this decay has fundamental applications
in all branches of genetics, as it measures how studies generalise to different
populations.

\section*{Introduction}

Predicting unobserved phenotypes using high-density SNP or sequence data
is the foundation of many applications in medical diagnostics
\cite{prenatal,cancer,celiac}, plant \cite{alison13,spindel} and animal 
\cite{goddard} breeding. The accuracy of genomic predictions will depend
on a number of factors: relatedness among genotyped individuals
\cite{goddard2,balding}; the density of the markers
\cite{visscher,frank,goddard2}; and the genetic architecture of the trait,
in particular the allele frequencies of causal variants \cite{rare1,rare2}
and the distribution of their effect sizes \cite{goddard2}.

Most of these issues have been explored in the literature, and have been 
tackled in various ways either from a methodological perspective or by
producing larger data sets and more accurate phenotyping. However, the extent
to which predictive models generalise from the populations used to train them
to distantly related target populations appears not to have been widely
investigated (two exceptions are \cite{goddard2,reviewer}). The accuracy of
prediction models is often evaluated in a general setting using cross-validation
with random splits, which implicitly assumes that test individuals are drawn
from the same population as the training sample; in that case accuracy to
predict phenotypes is only bounded by heritability, although unaccounted
``missing heritability'' is common \cite{makowsky,campos}. However, this
assumption is violated in many practical applications, such as genomic
selection, that require predictions of individuals that are genetically
distinct from the training sample: for instance, causal variants may differ in
both frequency and effect size between different ancestry groups (in humans,
e.g. \cite{lactose} for lactose persistence), subspecies (in plants and animals,
e.g. \cite{rice} for rice) or even families \cite{hickey2}. In such cases
cross-validation with random splits may overestimate predictive accuracy due to
the mismatch between model validation and the prediction problem of interest
\cite{hickey,habier} even when population structure is taken into account
\cite{deroos}. The more distantly the target population is related to the
training population, the lower the average predictive accuracy of a genomic
model; this has been demonstrated on both simulated and real dairy cattle data 
\cite{habier,pszczola,clark}. 

In this paper we will investigate the relationship between genetic distance and
predictive accuracy in the prediction of quantitative traits. We will simulate
training and target samples with varying genetic distances by splitting the
training population into a sequence of pairs of subsets with increasing genetic
differentiation. We will measure predictive accuracy with Pearson's correlation,
which we will estimate by performing genomic prediction from one subset to the
other in each pair. Among various measures of relatedness available in the
literature, we will consider mean kinship and $\F$, although we will only focus
on the latter. We will then study the mean Pearson's correlation as a
function of genetic distance, which we will refer to as the ``decay curve''
of the former over the latter. 

This approach is valuable in addressing several key questions in the
implementation of genomic selection programs, such as: How often (e.g., in terms
of future generations) will the genomic prediction model have to be re-estimated
to maintain a minimum required accuracy in the predictions of the phenotypes?
How should we structure our training population to maximise that accuracy? Which
new, distantly related individuals would be beneficial to introduce in a 
selection program for the purpose of maintaining a sufficient level of genetic
variability?

\section*{Materials and Methods}

\subsection*{Genomic Prediction Models}

A baseline model for genomic prediction of quantitative traits is the 
genomic BLUP (GBLUP; \cite{vanraden,meuwissen}), which is usually written as
\begin{align}
  \label{eq:blup}
  &\mathbf{y} = \boldsymbol{\mu} + \mathbf{Zg} + \boldsymbol{\varepsilon}&
  &\text{with}&
  &\mathbf{g} \sim N(\mathbf{0}, \mathbf{K}\sigma^2_g)
  &\text{and}&
  &\boldsymbol{\varepsilon} \sim N(\mathbf{0}, \sigma^2_\varepsilon),
\end{align}
where $\mathbf{g}$ is a vector of genetic random effects, $\mathbf{Z}$ is a
design matrix that can be used to indicate the same genotype exposed to
different environments, $\mathbf{K}$ is a kinship matrix and 
$\boldsymbol{\varepsilon}$ is the error term. Many of its properties are
available in closed form thanks to its simple definition and normality
assumptions, including closed form expressions of and upper bounds on
predictive accuracy that take into account possible model misspecification
\cite{campos}. Other common choices are additive linear regression models of
the form
\begin{equation}
  \label{eq:bayesreg}
  \mathbf{y} = \boldsymbol{\mu} + \mathbf{X}\boldsymbol{\beta} + 
    \boldsymbol{\varepsilon}
\end{equation}
where $\mathbf{y}$ is the trait of interest; $\mathbf{X}$ are the markers (such
as SNP allele counts coded as $0$, $1$ and $2$ with $1$ the heterozygote); 
$\boldsymbol{\beta}$ are the marker effects; and $\boldsymbol{\varepsilon}$ are
independent, normally-distributed errors with variance $\sigma^2_{\varepsilon}$.
Depending on the choice of the prior distribution for $\boldsymbol{\beta}$, we
can obtain different models from the literature such as BayesA and BayesB 
\cite{meuwissen}, ridge regression \cite{ridge}, the LASSO \cite{lasso} or the
elastic net \cite{elastic}. The model in Eq. (\ref{eq:blup}) is equivalent to
that in Eq. (\ref{eq:bayesreg}) if the kinship matrix $\mathbf{K}$ is computed
from the markers $\mathbf{X}$ and has the form $\mathbf{XX}^T$ and 
$\boldsymbol{\beta} \sim N(0, \VAR(\boldsymbol{\beta}))$
\cite{rrblup2,stranden}. In the remainder of the paper we will focus on the
elastic net, which we have found to outperform other predictive models on
real-world data \cite{sagmb12}. This has been recently confirmed in
\cite{zivan}.

Predictive accuracy is often measured by the Pearson correlation ($\hat\rho$)
between the predicted and observed phenotypes. When we use the fitted values
from the training population as the predicted phenotypes, and assuming that the
model is correctly specified, $\hat\rho^2$ coincides with the proportion of
genetic variance of the trait explained by the model and therefore
$\hat\rho^2 \ls h^2$, the heritability of the trait. (An incorrect model may
lead to overfitting, and in that case $\hat\rho^2 \gs h^2$.) When using
cross-validation with random splits, $\rCV \ls \hat\rho$ and typically the
difference will be noticeable ($\rCV \ll \hat\rho$). However, $\hat\rho_{CV}$
may still overestimate the actual predictive accuracy $\rD$ in practical
applications where target individuals for prediction are more different from
the training population than the test samples generated using cross-validation
\cite{makowsky}. This problem may be addressed by the use of alternative model
validation schemes that mirror more closely the prediction task of interest;
for instance, by simulating progeny of the training population to assess
predictive accuracy for a genomic selection program. This approach is known
as forward prediction and is common in animal breeding \cite{hickey,perez}.

Another possible choice is the prediction error variance (PEV). It is commonly
used in conjunction with GBLUP because, for that model, it can be estimated
(for small samples) or approximated (for large samples) in closed form from
Henderson's mixed model equations \cite{tier}. In the general case no closed
form estimate is available, but PEV can still be derived from Pearson's
correlation \cite{meuwissen2} for any kind of model as both carry the same
information:
\begin{equation}
  \mathrm{PEV} = (1 - \hat\rho^2) * \VAR(\mathbf{y}).
\end{equation}
For consistency with our previous work \cite{sagmb12} and with \cite{alison13},
whose results we partially replicate below, we will only consider predictive
correlation in the following. 

\subsection*{Kinship Coefficients and $\F$}

A common measure of kinship from marker data is average allelic correlation 
\cite{vanraden,astle}, which is defined as $\mathbf{K} = [ k_{ij} ]$ with
\begin{equation}
\label{eq:allelic}
  k_{ij} = \frac{1}{m}\sum_{k = 1}^m \tilde{X}_{ik} \tilde{X}_{jk}
\end{equation}
where $\tilde{X}_{ik}$ and $\tilde{X}_{jk}$ are the standardised allele counts
for the $i$th and $j$th individuals and the $k$th marker. An important property
of allelic correlation is that it is inversely proportional to the Euclidean
distance between the marker profiles $X_i, X_j$ of the corresponding
individuals: if the markers are standardised
\begin{equation}
  \sqrt{2n - 2k_{ij}} = 
  \sqrt{2n - 2\sum_{k = 1}^m \tilde{X}_{ik}\tilde{X}_{jk}} =
  \sqrt{\sum_{k = 1}^m \tilde{X}_{ik}^2 + \tilde{X}_{jk}^2 - 2\tilde{X}_{ik}\tilde{X}_{jk}} =
  \sqrt{\sum_{k = 1}^m (\tilde{X}_{ik} - \tilde{X}_{jk})^2}.
\end{equation}

This result has been used in conjunction with clustering methods
such as \km{} or partitioning around medoids (PAM; \cite{bishop}) to produce
subsets of minimally related individuals from a given sample by maximising the
Euclidean distance \cite{saatchi,hickey,makowsky}. 

At the population level, the divergence between two populations due to drift,
environmental adaptation, or artificial selection is commonly measured with
$\F$. Several estimators are available in the literature, and reviewed in
\cite{bhatia}. In this paper we will adopt the estimator from \cite{balding2},
which is obtained by maximising the Beta-Binomial likelihood of the allele
frequencies as a function of $\F$. $\hF$ then describes how far the target
population has diverged from the training population, which translates to
``how far'' a genomic prediction model will be required to predict. In terms
of kinship, we know from the literature that the mean kinship coefficient
$\bk$ between two individuals in different populations is inversely related to
$\hF$ \cite{kimura}: kinship can be interpreted as the probability that two
alleles are identical by descent, which is inversely related to $\F$ which is a
mean inbreeding coefficient. Intuitively, the fact that individuals in the two
populations are closely related implies that the latter have not diverged much
from the former: if $\bk$ is large, the marker profiles (and therefore the
corresponding allele frequencies) will on average be similar. As a result, any
clustering method that uses the Euclidean distance to partition a population
into subsets will maximise their $\F$ by minimising $\bk$. The simulations
and data analyses below confirm experimentally that $\bk$ and $\hF$ are highly
correlated, which makes them equivalent in building the decay curves; thus we
will report results only for $\hF$ (see Section C, \nameref{S1_Text}).

\subsection*{Real-World Data Sets}

We evaluate our approach to construct decay curves for predictive accuracy using
two publicly-available real-world data sets with continuous phenotypic traits,
and a third, human, genotype data set.

\textbf{WHEAT.} We consider $376$ wheat varieties from the TriticeaeGenome
project, described in \cite{alison13}. Varieties collected from those registered
in France ($210$ varieties), Germany ($90$ varieties) and the UK ($75$
varieties) between 1946 and 2007 were genotyped using a combination of $2712$
predominantly DArT markers. Several traits were recorded; in this paper we will
focus on grain yield, height, flowering time, and grain protein content.
Genotype-environment interactions were accounted for by an incomplete block
design over trial fields in different countries, to prevent genomic prediction
being biased by the country of registration of each variety. As in
\cite{alison13}, we also group varieties in three groups based on their year of
registration: pre-1990 ($103$ varieties), 1990 to 1999 ($120$ varieties), and
post-1999 ($153$ varieties).

\textbf{MICE.} The heterogeneous mice population from \cite{mice2} consists of
$1940$ individuals genotyped with $12545$ SNPs; among the recorded traits, we
consider growth rate and weight. The data include a number of inbred families,
the largest being F005 ($287$ mice), F008 ($293$ mice), F010 ($332$ mice) and
F016 ($309$ mice).

\textbf{HUMAN.} The marker profiles from the Human Genetic Diversity Panel
\cite{hgdp-snp} include $1043$ individuals from different ancestry groups: $151$
from Africa, $108$ from America, $435$ from Asia, $167$ from Europe, $146$ from
the Middle East and $36$ from Oceania. Each has been genotyped with $650,000$
SNPs; for computational reasons we only use those in chromosomes $1$ and $2$,
for a total of $90,487$ SNPs.

All data sets have been pre-processed by removing markers with minor allele
frequencies $< 1\%$ and those with $> 20\%$ missing data. The missing data in
the remaining markers have been imputed using the \pkg{impute} R package
\cite{impute}. Finally, we removed one marker from each pair whose allele
counts have correlation $> 0.95$ to increase the numerical stability of the
genomic prediction models.

\subsection*{Decay Curves for Predictive Accuracy}

We estimate a decay curve of $\rD$ as a function of $\F$ as follows:
\begin{enumerate}
  \item Produce a pair of minimally related subsets (i.e., with maximum $\F$)
    from our training population using \km{} clustering, $k = 2$ in R
    \cite{rcore}. PAM was also considered as an alternative clustering method,
    but produced subsets identical to those from \km{} for all the data sets
    studied in this paper. The largest of these two subsets
    will be used to train the genomic prediction model, and will be considered
    the ancestral population for the purposes of computing $\F$; the smallest
    will be the target used for prediction. In the following we will call them
    the \textit{training subsample} and the \textit{target subsample},
    respectively.
    \label{item:kmeans}
  \item Compute $\hF^{(0)}$ and $\rD^{(0)}$ for the pair of subsets with a 
    genomic prediction model. We compute $\hF^{(0)}$ using the Beta-Binomial
    estimator from \cite{balding2}; and we compute $\rD^{(0)}$ with the elastic
    net implementation in the \pkg{glmnet} R package \cite{glmnet}. Other models
    can be used: the proposed approach is model-agnostic as it only requires the
    chosen model to be able to produce estimates of its predictive correlation.
    The optimal values for the penalty parameters of the elastic net are chosen
    to maximise $\rCV$ on the training subset using $5$ runs of $10$-fold
    cross-validation as in \cite{waldron}. $(\hF^{(0)}, \rD^{(0)})$ will act as
    the far end of the decay curve (in terms of genetic distance).
    \label{item:methods}
  \item For increasing numbers $m$ of individuals:
    \begin{enumerate}
      \item create a new pair of subsamples by swapping $m$ individuals at
        random between the training and the test subsamples from step
        \ref{item:kmeans};
      \item fit a genomic prediction model on the new training subsample and
        use it to predict the new target subsample, thus obtaining 
        $(\hF^{(m)}, \rD^{(m)})$ using the same algorithms as in step
        \ref{item:methods}.
    \end{enumerate}
  \item Estimate the decay curve from the sets of $(\hF^{(m)}, \rD^{(m)})$
    points using local regression (LOESS; \cite{loess}), which can be used to
    produce both the mean and its $95\%$ confidence interval at any point in
    the range of observed $\hF$. We denote with $\rD$ the resulting estimate
    of predictive correlation for any given $\hF$.
\end{enumerate}
The pair of subsets produced by \km{} corresponds to $m = 0$, hence the notation
$(\hF^{(0)}, \rD^{(0)})$, and we increase $m$ by steps of $2$ to $20$ until the
$\hF$ between the subsamples is at most $0.005$. We choose the stepping for each
data set to be sufficiently small to cover the interval $[0, \hF^{(0)}]$ as
uniformly as possible. The larger $m$ is, the smaller we can expect $\hF^{(m)}$
to be. We repeat step 3(a) and 3(b) $40$ times for each $m$ to achieve the 
precision needed for an acceptably smooth curve.

As an alternative approach, we also consider estimating the decay rate of $\rD$
by linear regression of the $\rD^{(m)}$ against the $\hF^{(m)}$; we will denote
the resulting predictive accuracy estimates with $\rL$. For any set value of
$\hF$, we compare the $\rL$ at that $\hF$ with the corresponding value $\rD$
from the decay curve estimated by averaging all the $\rD^{(m)}$ for which
$|\hF^{(m)} - \hF| \ls 0.01$. Assuming that the decay curve is in fact a
straight line reduces the number of subsamples that we need to generate,
enforces smoothness and makes it possible to compute $\rL$ for values of $\F$
larger than $\hF^{(0)}$. On the other hand, the estimated $\rL$ will be
increasingly unreliable as $\rL \to 0$, because the regression line will provide
negative $\rL$ instead of converging asymptotically to zero. We also regress the
$\left(\rD^{(m)}\right)^2$ against the $\left(\hF^{(m)}\right)^2$ to
investigate whether they have a stronger linear relationship than the
$\rD^{(m)}$ with the $\hF^{(m)}$, as suggested in \cite{pszczola} using
simulated genotypes and phenotypes mimicking a dairy cattle population.

The size of the training ($\nTR$) and target ($\nTA$) subsamples is determined
by \km{}. For the data used in this paper, \km{} splits the training populations
in two subsamples of comparable size; but we may require a smaller $\nTA 
\ll \nTR$ to estimate $\rD^{(0)}$ and the $\rD^{(m)}$ while at the same time a
larger $\nTR$ is needed to fit the genomic prediction model. In that case, we
increase $\nTR$ by moving individuals from the target subsample while keeping
the $\hF^{(0)}$ between the two as large as possible. The impact on the
estimated $\hF$ is likely to be small, because its precision depends more on
the number of markers than on $\nTR$ and  $\nTA$ \cite{balding2}. The estimated
$\rD^{0}$ and $\rD^{(m)}$ might be inflated because we are altering the subsets,
even when $\hF$ does not change appreciably. Its variance, which can be 
approximated as in \cite{hooper}, decreases linearly in $\nTA$ except that can
be compensated by generating more pairs of subsamples for each value of $m$.

\subsection*{Simulation Studies}

We study the behaviour of the decay curves via two simulation studies. 

\textbf{Genomic selection.} We simulate a genomic selection program using the
wheat varieties registered in the last $5$ years of the WHEAT data as founders.
The simulation is a forward simulation implemented as follows for $10$, $50$,
$200$ and $1000$ causal variants, and decay curves are produced for each.

\begin{enumerate}
  \item We set up a training population of $200$ founders: $96$ varieties from
    the WHEAT data, $104$ obtained from the former via random mating without
    selfing using the \textbf{HaploSim} R package \cite{haplosim}.
    \textbf{HaploSim} assumes that markers are allocated at regular intervals
    across the genome, we allocated them uniformly in $21$ chromosomes (wheat
    varieties in the WHEAT data are allohexaploid, with $2n$ = $6x$ = $42$) to
    obtain roughly the desired amount of recombination and to preserve the
    linkage disequilibrium patterns as much as possible. \label{item:geno}
  \item We generate phenotypes by selecting causal variants at random among
    markers with minor allele frequency $> 5\%$ and assigning them 
    normally-distributed additive effects with mean zero. Noise is likewise
    normally distributed with mean zero and standard deviation $1$, and the
    standard deviation of the additive effects is set such that
    $h^2 \approx 0.55$. We choose this value as the mid-point of a range of
    heritabilities, $[0.40, 0.70]$, we consider to be of interest.
\label{item:pheno}
  \item We fit a genomic prediction model on the whole training population.
    \label{item:gs}
  \item For $100$ times, we perform a sequence of $10$ rounds of selection. In
    each round:
  \begin{enumerate}
    \item we generate the marker profiles of $200$ progeny via random mating,
      again without selfing; \label{item:markers}
    \item we generate the phenotypes for the progeny as in step
      \ref{item:pheno}; \label{item:curpheno}
    \item we compute the $\hF$ between the training population and the progeny
      generated in \ref{item:markers};
    \item we use the marker profiles from step \ref{item:markers} and the
      genomic prediction model from \ref{item:gs} to obtain predicted values
      for the phenotypes, which are then used together with those from step
      \ref{item:curpheno} to compute predictive correlation;
    \item we select the $20$ individuals with the largest phenotypes as the
      parents of the next round of selection.
  \end{enumerate}
  \item We compute the average predictive correlation $\rb$ and the average
    $\hF$ for each round of selection, which are used as reference points
    to assess how well the results of the genomic selection simulation are
    predicted by the decay curve. \label{item:avg}
  \item We estimate the decay curve $(\hF^{(m)}, \rD^{(m)})$ and its linear
     approximation $\rL$ from the training population, and we compare it with
     the average $(\hF, \rb)$ reference points from step \ref{item:avg}.
\end{enumerate}

We then repeat this simulation after adding the varieties available at the end
of the second round of selection to the training population while considering
the scenario with $200$ and $1000$ causal variants. The size of the training
population is thus increased to $800$ varieties, allowing us to explore the
effects of a larger sample size and of considering new varieties from the
breeding program to update the genomic prediction models when their predictive
accuracy is no longer acceptable. In the following, we refer to this second
population as the ``augmented population'' as opposed to the ``original
population'' including only the $200$ varieties described in steps 
\ref{item:geno} and \ref{item:pheno} above.

\textbf{Cross-population prediction.} We explore cross-population predictions
using the HUMAN data and simulated phenotypes. Similarly to the above, we pick
$5$, $20$, $100$, $2000$, $10000$ and $50000$ causal variants at random among
those with minor allele frequency $> 5\%$ and we assign them
normally-distributed effects such that $h^2 \approx 0.55$. The same effect sizes
are used for all populations. We then use individuals from Asia as the training
population to estimate the decay curves. Those from other continents are the
target populations for which we are assessing predictive accuracy, and we
compute their $\hF$ and the corresponding predictive correlations $\rP$. We use
the $(\hF, \rP)$ points as terms of comparison to assess the quality of the
curve, which should be close to them or at least cross the respective $95\%$
confidence intervals. 

\subsection*{Real-World Data Analyses}

Finally, we estimate the decay curves for some of the phenotypes available in
the WHEAT and MICE data. For both data sets we also produce and average $40$
values of $\rCV$ using hold-out cross-validation. In hold-out cross-validation
we repeatedly split the data at random into training and target subsamples
whose sizes are fixed to be the same as those arising from clustering in step
\ref{item:kmeans} of the decay curve estimation. Then we fit an elastic net
model on the training subsamples and predict the phenotypes in the target
subsamples to estimates $\rCV$. Ideally, the decay curve should cross the area
in which the $(\hF, \rCV)$ points cluster.

\textbf{WHEAT data.} For the WHEAT data, we construct decay curves for grain
yield, height, flowering time and grain protein content using the French wheat
varieties as the training population. UK and German varieties are the target
populations, for which we estimate $(\hF, \rP)$. Furthermore, we also construct
a second decay curve for yield using the varieties registered before 1990 as
the training population, as in \cite{alison13}. Varieties registered between
1990 and 1999, and those registered after 2000, are used as target populations.

\textbf{MICE data.}
For the MICE data, we construct decay curves for both growth rate and weight
using each of the F005, F008, F010 and F016 inbred families in turn as the 
training population; the remaining families are used as target populations. 

\section*{Results}

\subsection*{General Considerations}

\begin{table}[ht!]
%\begin{adjustwidth}{-2.25in}{0in}
  \caption{\bf Summary of the predictive correlations defined in the Methods.}
  \begin{tabular}{|c|p{12cm}|}
  \hline
  $\rCV$ & Predictive correlation computed on the whole training population by
           hold-out cross-validation with random splits. \\
  \hline
  $\rD^{(m)}$ & Predictive correlation for a target subsample computed from a
                genomic prediction model fitted on the corresponding training
                subsample after swapping $m$ individuals between the two. Used
                to construct the decay curve via LOESS together with the
                corresponding $\hF^{(m)}$. The subsamples are created from the
                training population via clustering to be minimally related. \\
  \hline 
  $\rD$ & Predictive correlation estimated by the decay curve at a given $\hF$. \\
  \hline 
  $\rL$ & Linear approximation to the decay curve computed by regressing the
          $\rD^{(m)}$ against the associated $\hF^{(m)}$.\\
  \hline 
  $\rP$ & Predictive correlation for a target population computed by fitting
          a genomic prediction model on the whole training population, used as
          a reference point in assessing the decay curve. \\
  \hline 
  $\rb$ & Mean predictive correlation for a generation in the genomic selection
          simulation, computed from a genomic prediction model fitted on the
          founders. \\
  \hline
  \end{tabular}
  \label{tab:rhos}
\end{table}

\begin{figure}[h]
  \caption{\bf Simulation of a $10$-generation breeding program using $200$ 
    varieties from the WHEAT data.}
  \begin{center}
    \includegraphics[width=0.65\textwidth]{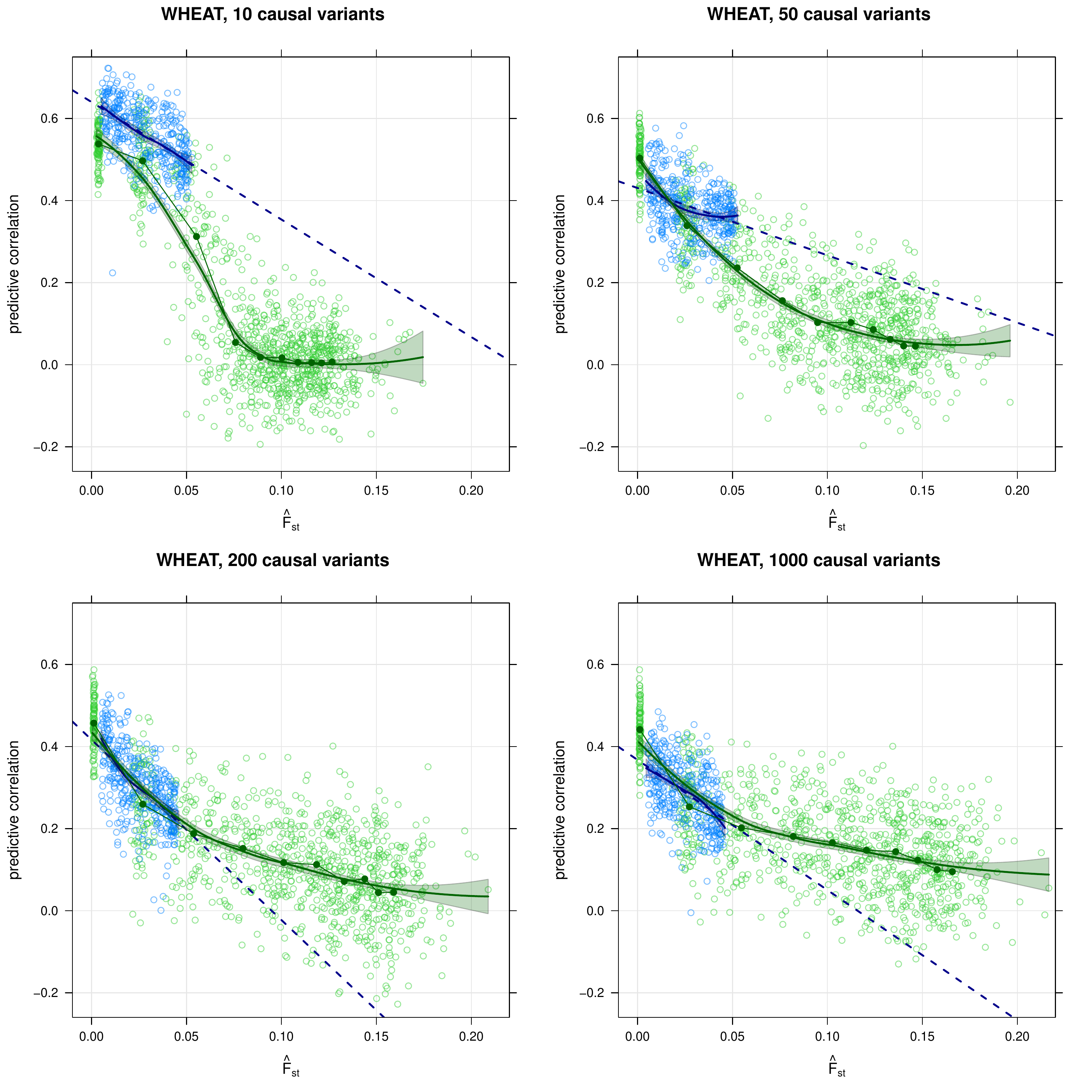}
  \end{center}
  Simulation of a $10$-generation breeding program developed using $200$
  varieties generated from $2002$--$2007$ WHEAT data with $10$ (top left),
  $50$ (top right), $200$ (bottom left) and $1000$ (bottom right) causal
  variants. The decay curves, the $\rD^{(0)}$ and the $\rD^{(m)}$ are
  in blue, and their linear interpolation ($\rL$) is shown as a dashed blue
  line. The open green circles are predictive correlations for the simulated
  populations, and the green solid points are the mean $(\hF, \rb)$ for
  each generation.
  \label{fig:breeding}
\end{figure}

The decay curves from the simulations are shown in Figs. \ref{fig:breeding},
\ref{fig:breeding2} and \ref{fig:hgdp}, and the corresponding predictive
correlations are reported in Tables 1 and 2, \nameref{S1_Text}. The predictive
correlations for the WHEAT and MICE data sets are reported in Table
\ref{tab:refpoints}, and the decay curves are shown in Figs. 1, 2 and 3,
\nameref{S1_Text}. A summary of the different predictive correlations defined
in the Methods and discussed here is provided in Table \ref{tab:rhos}.

In all the simulations and the real-world data analyses the $\rD$ from the
decay curve is close to the linear interpolation $\rL$; considering all the
reference populations in Table \ref{tab:refpoints} and the generation means in
Tables A.1 and A.2, \nameref{S1_Text}, $|\rD - \rL| \ll 0.02$ $41$ times out
of $47$ ($87\%$). Both estimates of predictive correlation are close to the
respective reference values $\rb$ and $\rP$; the difference (in absolute value)
is $\ll 0.05$ $39$ times ($41\%$) and $\ll 0.10$ $69$ times ($73\%$) out of
$94$. The proportion of small differences increases when considering only
target populations that fall within the span of the decay curve: $23$ out of
$44$ ($52\%$) are $\ll 0.05$ and $38$ are $\ll 0.10$ ($84\%$). This is
expected because the decay curve is already an extrapolation from the training
population, so extending it further with the linear interpolation $\rL$ 
reduces its precision. Regressing $\left(\rD^{(m)}\right)^2$ against the
$\left(\hF^{(m)}\right)^2$ does not produce a stronger linear relationship
than that represented by $\rL$ ($p = 0.784$, see Section D, \nameref{S1_Text}).

The range of the predictive correlations $\rD^{(m)}$ around the decay curves
varies between $0.05$ and $0.10$, and it is constant over the range of observed
$\hF$ for each curve. It does not appear to be related to either the size of
the training subsample or the number of causal variants. This is apparent in
particular from the genomic selection simulation, in which both are jointly
set to different combinations of values. Similarly, there seems to be no
relationship between the spread and the magnitude of the predictive correlations
($\rD^{(m)} \in [0, 0.75]$). This amount of variability is comparable to that of
other studies (e.g., the range of the $\rD^{(m)}$ is smaller than that in the
cross-validated correlations in \cite{zivan}) once we take into account that the
$(\hF^{(m)}, \rD^{(m)})$ are individual predictions and are not averaged over
multiple repetitions. Furthermore, subsampling further reduces the size of the
training subpopulations; and fitting the elastic net requires a search over a
grid of values for its two tuning parameters, which may get stuck in local
optima.

\begin{table}[ht!]
%\begin{adjustwidth}{-2.25in}{0in}
  \caption{\bf Predictive correlations for the analyses shown in 
    Figures B.1, B.2 and B.3, \nameref{S1_Text}.}
  \begin{tabular}{|p{3cm}|p{2cm}|p{2cm}|r|r|c|c|c|c|}
  \hline
  \textbf{Trait} & \textbf{Training \newline Population} & \textbf{Target \newline Population} & 
    $\nTR$ & $\nTA$ & $\hF^{(0)}$ & $\rP$ & $\rD$ & $\rL$ \\
  \hline
  \multirow{2}{*}{WHEAT, Yield}
    & France & UK      & $132$ & $70$ & $0.031$ & $0.55$ & $0.60$ & $0.58$ \\
  \cline{2-9}
    & France & Germany & $132$ & $70$ & $0.042$ & $0.56$ & $0.56$ & $0.51$ \\
  \hline
  \multirow{2}{*}{WHEAT, Height}
    & France & UK      & $132$ & $70$ & $0.031$ & $0.57$ & $0.63$ & $0.58$ \\
  \cline{2-9}
    & France & Germany & $132$ & $70$ & $0.042$ & $0.60$ & $0.55$ & $0.54$ \\
  \hline
  \multirow{2}{*}{\parbox{3cm}{WHEAT, \newline Flowering time}}
    & France & UK      & $132$ & $70$ & $0.031$ & $0.36$ & $0.70$ & $0.70$ \\
  \cline{2-9}
    & France & Germany & $132$ & $70$ & $0.042$ & $0.23$ & $0.67$ & $0.68$ \\
  \hline
  \multirow{2}{*}{\parbox{3cm}{WHEAT, Grain \newline protein content}}
    & France & UK      & $132$ & $70$ & $0.031$ & $0.59$ & $0.54$ & $0.51$ \\
  \cline{2-9}
    & France & Germany & $132$ & $70$ & $0.042$ & $0.47$ & $0.46$ & $0.45$ \\
  \hline
  \multirow{12}{*}{MICE, Weight}
    & F005 & F008 & $155$ & $132$ & $0.065$ & $0.14$ & $0.18$ & $0.21$ \\
  \cline{2-9}
    & F005 & F010 & $155$ & $132$ & $0.062$ & $0.17$ & $0.20$ & $0.21$ \\
  \cline{2-9}
    & F005 & F016 & $155$ & $132$ & $0.061$ & $0.15$ & $0.20$ & $0.22$ \\
  \cline{2-9}
    & F008 & F005 & $203$ & $90^*$ & $0.066$ & $0.24$ & -    & $0.30$ \\
  \cline{2-9}
    & F008 & F010 & $203$ & $90^*$ & $0.063$ & $0.21$ & -    & $0.31$ \\
  \cline{2-9}
    & F008 & F016 & $203$ & $90^*$ & $0.056$ & $0.16$ & -    & $0.34$ \\
  \cline{2-9}
    & F010 & F005 & $241$ & $90^*$ & $0.063$ & $0.39$ & -    & $0.52$ \\
  \cline{2-9}
    & F010 & F008 & $241$ & $90^*$ & $0.062$ & $0.22$ & -    & $0.52$ \\
  \cline{2-9}
    & F010 & F016 & $241$ & $90^*$ & $0.067$ & $0.18$ & -    & $0.52$ \\
  \cline{2-9}
    & F016 & F005 & $238$ & $70^*$ & $0.063$ & $0.34$ & $0.29$ & $0.35$ \\
  \cline{2-9}
    & F016 & F008 & $238$ & $70^*$ & $0.057$ & $0.07$ & $0.32$ & $0.35$ \\
  \cline{2-9}
    & F016 & F010 & $238$ & $70^*$ & $0.069$ & $0.27$ & -    & $0.30$ \\
  \hline
  \multirow{12}{*}{\parbox{3cm}{MICE, \newline Growth rate}}
    & F005 & F008 & $207$ & $80^*$ & $0.065$ & $0.10$ & $0.19$ & $0.20$ \\
  \cline{2-9}
    & F005 & F010 & $207$ & $80^*$ & $0.062$ & $0.02$ & $0.19$ & $0.20$ \\
  \cline{2-9}
    & F005 & F016 & $207$ & $80^*$ & $0.061$ & $0.05$ & $0.20$ & $0.20$ \\
  \cline{2-9}
    & F008 & F005 & $199$ & $90^*$ & $0.066$ & $0.18$ & -    & $0.19$ \\
  \cline{2-9}
    & F008 & F010 & $199$ & $90^*$ & $0.063$ & $0.08$ & -    & $0.19$ \\
  \cline{2-9}
    & F008 & F016 & $199$ & $90^*$ & $0.056$ & $0.05$ & -    & $0.21$ \\
  \cline{2-9}
    & F010 & F005 & $237$ & $90^*$ & $0.063$ & $0.03$ & $0.12$ & $0.13$ \\
  \cline{2-9}
    & F010 & F008 & $237$ & $90^*$ & $0.062$ & $0.07$ & $0.12$ & $0.14$ \\
  \cline{2-9}
    & F010 & F016 & $237$ & $90^*$ & $0.067$ & $0.01$ & -    & $0.11$ \\
  \cline{2-9}
    & F016 & F005 & $219$ & $90^*$ & $0.063$ & $0.00$ & -    & $0.05$ \\
  \cline{2-9}
    & F016 & F008 & $219$ & $90^*$ & $0.057$ & $0.06$ & $0.07$ & $0.06$ \\
  \cline{2-9}
    & F016 & F010 & $219$ & $90^*$ & $0.069$ & $0.04$ & -    & $0.03$ \\
  \hline
  \end{tabular}
  \begin{flushleft}
  $\rP$ is the predictive correlation for the target population from the full
  training population. $\rD$ is the decay curve estimate of $\rP$, and is only
  available if the target population falls within the span of the decay curve.
  $\rL$ is the corresponding estimate from the linear extrapolation. $\nTR$ is
  the size of the training subsamples and $\nTA$ is the size of the target
  subsamples; those marked with an asterisk have been reduced to increase
  $\nTR$.
  \end{flushleft}
  \label{tab:refpoints}
  %\end{adjustwidth}
\end{table}

\subsection*{Real-World Data Analyses}

Several interesting points arise from the analysis of the real phenotypes in
the WHEAT and MICE data, shown in Table \ref{tab:refpoints} and in Figures B.1,
B.2 and B.3, \nameref{S1_Text}. Firstly, cross-validation always produces pairs
of subsamples with $\hF \ls 0.01$ and high $\rCV$ that are located at the left
end of the decay curve. The average $\hF$ is $0.006$ for the WHEAT data and
$0.001$ for the MICE data, and the difference between the average $\rCV$ and the
corresponding $\rD$ is $\ll 0.02$ $10$ times out of $12$ ($83\%$, see Table
B.4, \nameref{S1_Text}). The spread of the $\rCV$ is also similar to that of
the $\rD^{(m)}$. Secondly, we note that in the WHEAT data all decay curves but
that for flowering time cross the $95\%$ confidence intervals for the
cross-country predictive correlations $\rP$ for Germany and UK reported in
\cite{alison13}. Even in the MICE data, in which all families are near the end
or beyond the reach of the decay curves, the latter (or their linear
approximations) cross the $95\%$ confidence intervals for the $\rP$ $18$ times
out of $24$ ($75\%$). However, we also note that those intervals are wide due to
the limited sizes of those populations.

Furthermore, the decay curves for the phenotypes in the WHEAT data confirm two
additional considerations originally made in \cite{alison13}. Firstly,
\cite{alison13} noted that the distribution of the Ppd-D1a gene, which is a
major driver of this flowering time, varies substantially with the country of
registration and thus cross-country predictions are not reliable. Figure B.1,
\nameref{S1_Text} shows that the decay curve vastly overestimates the
predictive correlation for both Germany and the UK. Splitting the WHEAT data
in two halves that contain equal proportions of both alleles of Ppd-D1a and
that are genetically closer overall ($\hF = 0.04$), we obtain a decay curve
that fits the predictive correlations reported in the original paper 
($\rD = 0.77$, $\rP = 0.79$). Secondly, we also split the data according to
their year of registration and use the oldest varieties (pre-1990) as a
training sample for predicting yield. Again the decay curve crosses the $95\%$
confidence intervals for the predictive correlations reported in \cite{alison13}
and the correlations themselves are within $0.05$ of the average $\rD$ from the
decay curve both for 1990-1999 ($\hF = 0.028$, $\rD = 0.44$, $\rP = 0.40$) 
and post-2000 ($\hF = 0.033$, $\rD = 0.44$, $\rP = 0.42$) varieties.

\subsection*{Simulation Studies}

The decay curves from the genomic selection simulation on the original training
population ($200$ varieties), shown in blue in Fig. \ref{fig:breeding}, span
two rounds of selection and three generations. When considering $200$ or $1000$
causal variants, the curve overlaps the mean behaviour of the simulated data
points (shown in green) almost perfectly: the difference between the generation
means $\rb$ and the decay curve is $\ls 0.06$ for the first three generations,
with the exception of the first generation in the simulation with $1000$ variants
($|\rb - \rD| = 0.09$). As the number of causal variants decreases ($50$, $10$),
the decay curve increasingly overestimates $\rb$, although the difference
remains $\ls 0.10$ for the first two generations; and both show a slower decay
than the $\rb$. This appears to be due to a few alleles of large effect becoming
fixed by the selection, leading to a rapid decrease of $\rb$ without a
corresponding rapid increase in $\hF$.

\begin{figure}[h]
  \caption{\bf Simulation of a $10$-generation breeding program with a 
    training population augmented to $800$ varieties, after two rounds of
    selection.}
  \begin{center}
    \includegraphics[width=0.65\textwidth]{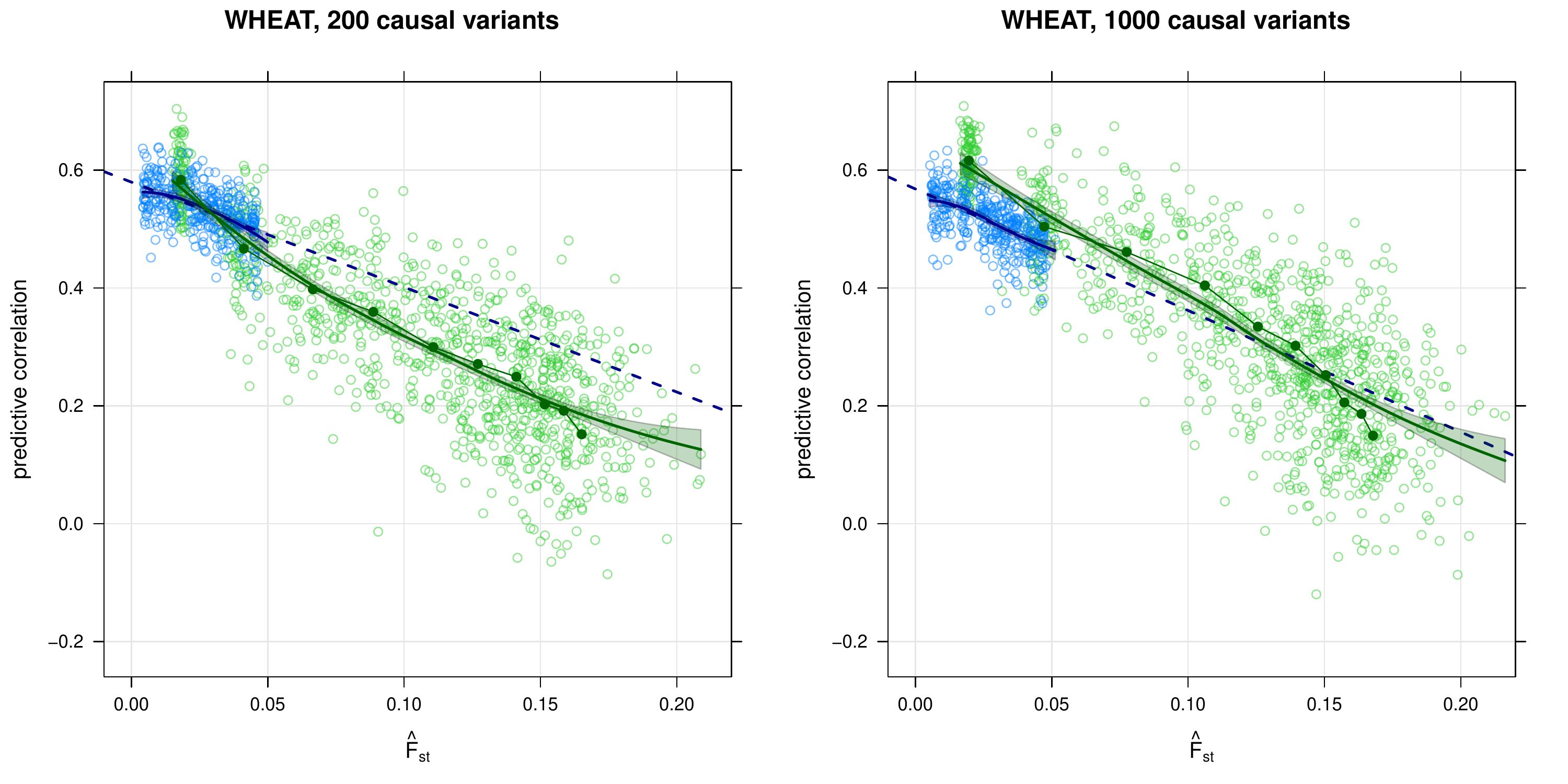}
  \end{center}
  Simulation of a $10$-generation breeding program with an updated genomic
  prediction model. The updated model is fitted on the $800$ varieties 
  available after the second round of selection in the simulations for $200$
  (left) and $1000$ (right) causal variants in Fig. \ref{fig:breeding}.
  Formatting is the same as in Fig. \ref{fig:breeding}.
  \label{fig:breeding2}
\end{figure}

The decay curves fitted on the augmented training populations ($800$ varieties,
now including those available at the end of the second round of selection, Fig.
\ref{fig:breeding2}) fit the first four generations well ($|\rb - \rD| \ls 0.04$
for the first two, $|\rb - \rD| \ls 0.06$ for the third and the fourth). As 
before, the only exception is the first generation in the simulation with $1000$
variants, with an absolute difference of $0.09$. However, the decay curves are
also able to capture the long-range decay rates through their linear
approximations. When considering $200$ causal variants, $|\rb - \rL| \approx 0.08$
for generations $5$ to $7$ and $\approx 0.10$ for generations $8$ and $9$; and
$|\rb - \rL| \ll 0.05$ for generations $4$ to $9$ when considering $1000$ causal
variants. This can be attributed to the increased sample size of the training
population, which both improves the goodness of fit of the estimated decay curve;
and makes the decay rate of the $\rb$ closer to linear, thus making it possible
for the $\rL$ to approximate it well over a large range of $\F$ values. To
investigate this phenomenon, we gradually increased the initial training
population to $4000$ varieties through random mating and we observed that for
such a large sample size $\rb$ indeed decreases linearly as a function of $\F$.
We conjecture that this is due to a combination of the higher values observed
for $\rb$ and their slower rate of decay, which prevents the latter from
gradually decreasing as $\rb$ is still far from zero after $10$ generations.
In addition, we note that increasing the number of causal variants has a similar
effect; with $200$ and $1000$ causal variants $\rb$ indeed decreases with an
approximately linear trend, which is not the case with $10$ and $50$ causal
variants.

\begin{figure}[p]
  \caption{\bf Simulation of quantitative traits from the HUMAN data.}
  \begin{center}
    \includegraphics[width=0.75\textwidth]{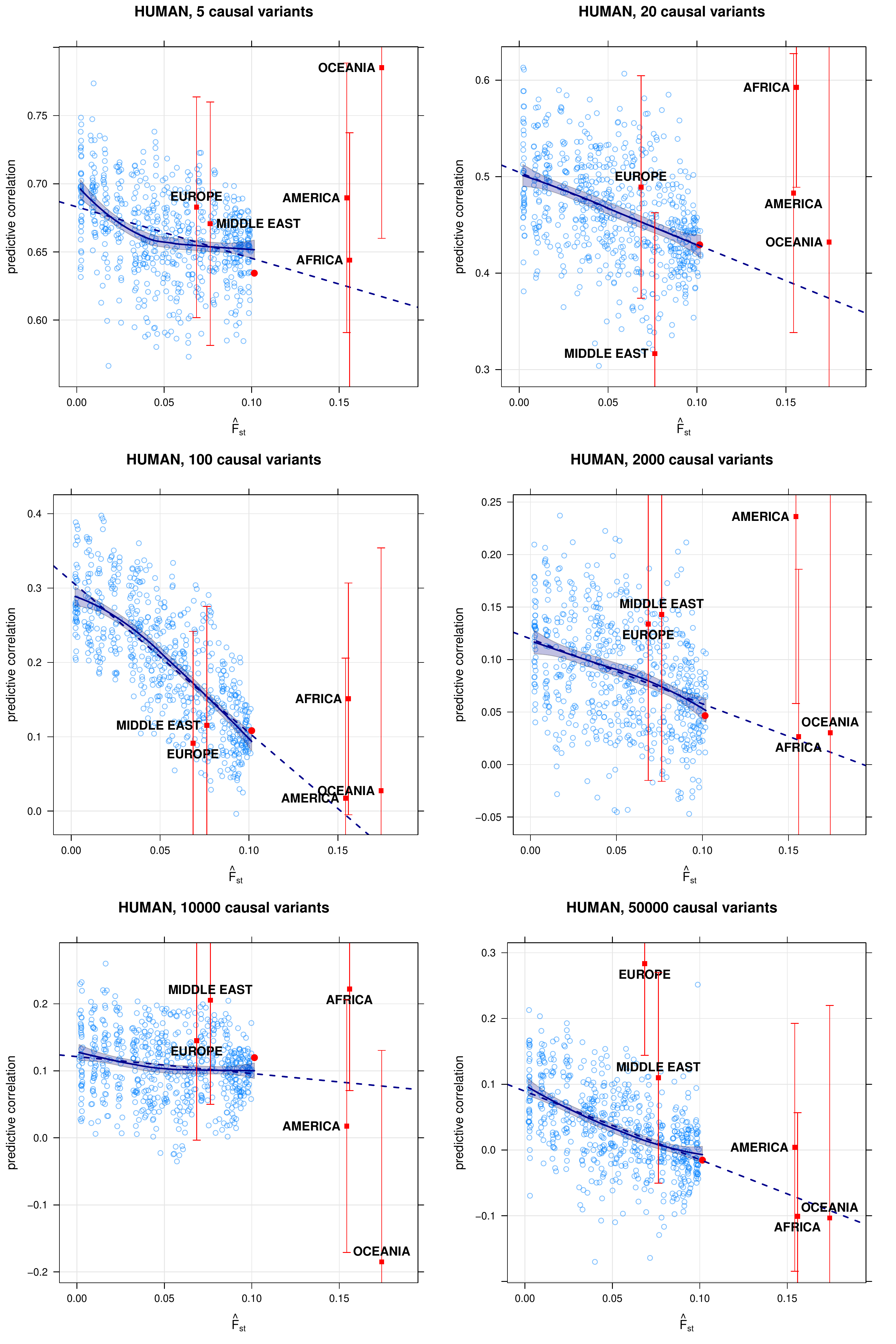}
  \end{center}
  \begin{flushleft}
  Simulation of quantitative traits with $5$ (top left), $20$ (top  right),
  $100$ (middle left), $2000$ (middle right), $10000$ (bottom left) and $50000$
  (bottom right) causal variants from the Asian individuals in the HUMAN data.
  The blue circles are the $\rD^{(m)}$ used to build the curve, and the red
  point is $\rD^{(0)}$. The blue line is the mean decay trend, with a shaded
  $95\%$ confidence interval, and the dashed blue line is the linear
  interpolation provided by the $\rL$. The red squares labelled EUROPE, MIDDLE
  EAST, AMERICA, AFRICA and OCEANIA correspond to the $\rP$ for the individuals
  from those continents, and the red brackets are the respective $95\%$
  confidence intervals.
  \end{flushleft}
  \label{fig:hgdp}
\end{figure}

The cross-population prediction simulation based on the HUMAN data (Fig.
\ref{fig:hgdp}) generated results consistent with those above. As before,
the number of causal variants appears to influence the behaviour of the decay
curve: while the $\rD^{(m)}$ decrease linearly for $20$, $100$ and $2000$
casual variants, they  converge to $0.65$ for $5$ causal variants. However,
unlike in the genomic selection simulation, the quality of the estimated
decay curve does not appear to degrade as the number of causal variants
decreases. This difference may depend on the lack of a systematic selection
pressure in the current simulation, which made the decay curve overestimate
predictive correlation when considering $10$ variants in the previous
simulation. Finally, as in the analysis of the MICE data, the linear 
approximation $\rL$ to the decay curve provides a way to extend the reach of
the decay curve to estimate predictive correlations $\rP$ for distantly related
populations (AMERICA, AFRICA, OCEANIA). Again we observe some loss in 
precision (see Table \ref{tab:refpoints}), but the extension still crosses the
$95\%$ confidence intervals of those $\rP$ $14$ times out of $18$ ($78\%$).

\section*{Discussion}

Being able to assess the predictive accuracy is important in many applications,
and will assist in the development of new models and in the choice of training
populations. A number of papers have discussed various aspects of the 
relationship between training and target populations in genomic prediction, and
of characterising predictive accuracy given some combination of genotypes and
pedigree information. For instance, \cite{rincent} discusses how to choose
which individuals to include in the training population to maximise prediction
accuracy for a given target population using the coefficient of determination.
\cite{habier3} separates the contributions of linkage disequilibrium,
co-segregation and additive genetic relationships to predictive accuracy,
which can help in setting expectations about the possible performance of
prediction. \cite{habier2} and \cite{pszczola} link predictive accuracy to
kinship in a simulation study of dairy cattle breeding; and \cite{lorenz} 
investigates the impact of population size, population structure and replication
in a simulated biparental maize populations. The approach we take in this paper
is different in a few, important ways. Firstly, we choose to avoid the
parametric assumptions underlying GBLUP and the corresponding approximations
based on Henderson's equations that provide closed-form results on predictive
accuracy in the literature. It has been noted in our previous work \cite{sagmb12}
and in the literature (e.g. \cite{zivan}) that in some settings GBLUP may not be
competitive for genomic prediction; hence we prefer to use models with better
predictive accuracy such as the elastic net for which the parametric assumptions
do not hold. Our model-agnostic approach is beneficial also because decay
curves can then be constructed for current and future competitive models, since
the only requirement of our approach is that they must be able to produce an
estimate of predictive correlation. Secondly, we demonstrate that the decay
curves estimated with the proposed approach are accurate in different settings
and on human, plant and animal real-world data sets. This complements previous
work that often used synthetic genotypes and analysed predictive accuracy in a
single domain, such as forward simulation studies on dairy cattle data.
Finally, we recognise that the target population whose phenotypes we would like
to predict may not be available or even known when training the model. In plant
and animal selection programs, one or more future rounds of crossings may not
yet have been performed; in human genetics, prediction may be required into
different demographic groups for which no training data are available.
Therefore, we are often limited to extrapolating a $\rD$ to estimate the $\rP$
we would observe if the target population were available. Prior information on
$\hF$ values is available for many species such as humans \cite{bhatia,hgdp-snp};
and can be used to extract the corresponding $\rD$ from a decay curve.

We observe that the decay rate of $\rD$ is approximately linear in $\hF$ for
most of the curves, suggesting that regressing the $\rD^{(m)}$ against the
$\hF^{(m)}$ is a viable estimation approach. This has the advantage of being
computationally cheaper than producing a smooth curve with LOESS since it
requires fewer $(\hF^{(m)}, \rD^{(m)})$ points and thus fewer genomic
prediction models to be fitted. In fact, if we assume that the decay rate is
linear we could also estimate it as the slope of the line passing through
$(\hF \approx 0, \rCV)$ and $(\hF^{(m)}, \rD^{(m)})$ for a single, small
value of $m$. It should be noted, however, that several factors can cause
departures from linearity, including the number of causal variants underlying
the trait, the use of small training populations and the confounding effect
of exogenous factors. In the case of the MICE data, for instance, predictions
may be influenced by cage effects; in the case of the WHEAT data, environmental
and seasonal effects might not be perfectly captured and removed by the trials'
experimental design. We also note that the decay curves for traits with small
heritabilities will almost never be linear, because $\rD$ converges
asymptotically to zero. Unlike the results reported in \cite{pszczola}, we do
not find a statistically significant difference between the strength of the
linear relationship between $\rD$ and $\hF$ and that between the respective
squares. There may be several reasons for this discrepancy; the simulation study
in \cite{pszczola} was markedly different from the analyses presented in this
paper, since it used simulated genotypes to generate the population structure
typical of dairy cattle and since it used GBLUP as a genomic prediction model.

We also observe that when $\hF^{(m)} \approx 0$, both $\rD^{(m)}$ and $\rL$ are,
as expected, similar to the $\rCV$ obtained by applying cross-validation to the
training populations selected from the WHEAT and MICE data. This suggests that
indeed $\rCV$ is an accurate measure of predictive accuracy only when the target
individuals for prediction are drawn from the same population as the training
sample, as previously argued by \cite{makowsky} and \cite{hickey}, among others.

Some limitations of the proposed approach are also apparent from the results
presented in the previous section. The most important of these limitations
appears to be that in the context of a breeding program the performance of the
decay curve depends on the polygenic nature of the trait being predicted, as
we can see by comparing the panels in Fig. \ref{fig:breeding}. This can be
explained by the fact that causal variants underlying less polygenic, highly
and moderately heritable traits will necessarily have some individually large
effects. As each of those variants approaches fixation due to selection
pressure, allele frequencies in key areas of the genome will depart from those
in the training population and the accuracy of any genomic prediction model will
rapidly decrease \cite{deroos}. However, these selection effects are genomically
local and so have little impact on $\hF$. A similar effect has been observed for
flowering time in the WHEAT data. \cite{alison13} notes that the Ppd-D1a gene is
a major driver of early flowering, but it is nearly monomorphic in one allele in
French wheat varieties and nearly monomorphic in the other allele in Germany and
the UK. As a result, even though the $\hF$ for those countries are as small as
$0.031$ and $0.042$, $\rD$ widely overestimates $\rP$ in both cases. A possible
solution would be to compute $\hF$ only on the relevant regions of the genome or,
if their precise location is unknown, on the relevant chromosomes; or to weight
$\hF$ to promote genomic regions of interest.

On the other hand, in the case of more polygenic traits a larger portion of the
genome will be in linkage disequilibrium with at least one causal variant, and
their effects will be individually small. Therefore, $\hF$ will increase more 
quickly in response to selection pressure and changes in predictive accuracy 
will be smoother, thus allowing $\rD$ to track them more easily. Indeed, in the
WHEAT data the genomic prediction model for flowering time has a much smaller
number of non-zero coefficients ($28$) compared to yield ($91$), height ($286$)
and grain protein content ($121$). Similarly, in the MICE data the model fitted
on F010 to predict weight has only $168$ non-zero coefficients while others range
from $212$ to $1169$ non-zero coefficients. By contrast, all models fitted for
predicting weight, which correspond to curves that well approximate other 
families' $\rP$, have between $1128$ and $2288$ non-zero coefficients.

The simulation on the HUMAN data suggests different considerations apply to 
outbred species. Having some large-effect causal variants does not necessarily
result in low quality decay curves; on the contrary, if we assume that the trait
is controlled by the same causal variants in the training and target populations
it is possible to have a good level of agreement between the $\rD$ and the $\rP$.
Intuitively, we expect strong effects to carry well across populations and thus
$\rD$ does not decrease beyond a certain $\F$. However, this will mean that the
curves will not be linear and $\rL$ will underestimate $\rP$ (see Fig.
\ref{fig:hgdp}, top left panel). We also note that effect sizes are the same
in all the populations, which may make our estimates of predictive accuracy
optimistic.

Another important consideration is that since the decay curve is extrapolated
from the training population, its precision decreases as $\F$ increases, as can
be seen from both simulations and by comparing the WHEAT and MICE data.
Predictions will be poor in practice if the target and the training populations
are too genetically distinct; an example are rice subspecies \cite{rice}, which
have been subject to intensive inbreeding. The trait to be predicted must have
a common genetic basis across training and target populations. However, the
availability of denser genomic data and of larger samples may improve both 
predictive accuracy and the precision of the decay curve for large $\F$. 
Furthermore, the range of the decay curve in terms of $\F$ depends on the amount
of genetic variability present in the training population; the more homogeneous
it is, the more unlikely that \km{} clustering will be able to split it in two
subsets with high $\hF^{(0)}$. One solution is to assume the decay is linear and
use $\rL$ instead of $\rD$ to estimate $\rP$; but as we noted above this is only
possible if $\rP \gg 0$. If $\rP \approx 0$, the decay curve estimated with
LOESS from $\rD$ can converge asymptotically to zero as $\hF$ increases; but the
linear regression used to estimate $\rL$ will continue to decrease until
$\rL \ll 0$. Another possible solution is to try to increase $\hF$ by moving
observations between the two subsets, but improvements are marginal at best and
there is a risk of inflating $\rD$.

Even with such limitations, estimating a decay curve for predictive correlation
has many possible uses. In the context of plant and animal breeding, it is a
useful tool to answer many key questions in planning genomic selection programs.
Firstly, different training populations (in terms of allele frequencies, sample
size, presence of different families, etc.) can be compared to choose that which
results in the slowest decay rate. Secondly, the decay curve can be used to
decide when genomic prediction can no longer be assumed to be accurate enough
for selection purposes, and thus how often the model should be re-trained on a
new set of phenotypes. Unlike genotyping costs, phenotyping costs for
productivity traits have not decreased over the years. Furthermore, the rate of
phenotypic improvements (i.e. selection cycle time) can be severely reduced by
the need of performing progeny tests. Therefore, limiting phenotyping to once
every few generations can reduce the cost and effort of running a breeding
program. The presence of close ancestors in the training population suggests
that decay curves are most likely reliable for this purpose, as we have shown
both in the simulations and in predicting newer wheat varieties from older ones
in the WHEAT data.

The other major application of decay curves is estimating the predictive
accuracy of a model for target populations that, while not direct descendants
of the training population, are assumed not to have strongly diverged and thus
to have comparable genetic architectures. Some examples of such settings are
the cross-country predictions for the WHEAT data, the cross-family predictions
for the MICE data and across human populations. In human genetics, decay curves
could be used to study the accuracy of predictions and help predict the success
of interventions of poorly-studied populations. In plant and animal breeding, on
the other hand, it is common to incorporate distantly related samples in 
selection programs to maintain a sufficient level of genetic variability. Decay
curves can provide an indication of how accurately the phenotypes for such
samples are estimated, since the model has not been trained to predict them
well and they are not as closely related as the individuals in the program.

\section*{Supporting Information}

\subsection*{S1 Text.}
\label{S1_Text}

\textbf{Supplementary information on the Methods and the Results.}
Figures for the decay curves from the simulation studies. Relationship
between $\hF$ and $\bk$. Comparison of the linear relationships between $\rho^2$
versus $\hF^2$ and $\rho$ and $\hF$.

\section*{Acknowledgements}

The work presented in this paper forms part of the MIDRIB project (``Molecular
Improvement of Disease Resistance in Barley''), which is funded by the UK
Technology Strategy Board (TSB) and Biotechnology \& Biological Sciences
Research Council (BBSRC), grant TS/I002170/1. The project was a collaboration
with Limagrain UK Ltd.; in particular, we would like to thank Anne-Marie 
Bochard and Mark Glew for their contributions. We would also like to thank
Jonathan Marchini (Department of Statistics, University of Oxford) and his
group for their insightful comments and suggestions.

\setcounter{figure}{0}
\setcounter{table}{0}
\renewcommand{\arraystretch}{1.2}
\renewcommand\thesection{\Alph{section}}
\renewcommand\thesubsection{\thesection.\arabic{subsection}}
\renewcommand{\thefigure}{\Alph{section}.\arabic{figure}}
\renewcommand{\thetable}{\Alph{section}.\arabic{table}}

\pagebreak

\section*{Supplementary Information}

\section{Simulation Studies}

\subsection{Breeding Program Simulation using the WHEAT data}

\begin{table}[h]
  \begin{center}
  \begin{tabular}{|c|c|c|c|c|c|}
  \hline
  \textbf{Causal Variants} & \textbf{Generation}
               & $\hF$   & $\rb$  & $\rD$  & $\rL$ \\
  \hline
  $10$   & $1$ & $0.003$ & $0.54$ & $0.61$ & $0.63$ \\
  \hline
  $10$   & $2$ & $0.027$ & $0.50$ & $0.56$ & $0.56$ \\
  \hline
  $10$   & $3$ & $0.055$ & $0.31$ & $0.49$ & $0.48$ \\
  \hline
  $50$   & $1$ & $0.001$ & $0.50$ & $0.44$ & $0.43$ \\
  \hline
  $50$   & $2$ & $0.026$ & $0.34$ & $0.38$ & $0.39$ \\
  \hline
  $50$   & $3$ & $0.052$ & $0.24$ & $0.36$ & $0.34$ \\
  \hline
  $200$  & $1$ & $0.001$ & $0.46$ & $0.40$ & $0.41$ \\
  \hline
  $200$  & $2$ & $0.027$ & $0.26$ & $0.29$ & $0.29$ \\
  \hline
  $200$  & $3$ & $0.053$ & $0.19$ & $0.23$ & $0.18$ \\
  \hline
  $1000$ & $1$ & $0.001$ & $0.44$ & $0.35$ & $0.36$ \\
  \hline
  $1000$ & $2$ & $0.027$ & $0.25$ & $0.29$ & $0.28$ \\
  \hline
  $1000$ & $3$ & $0.055$ & $0.20$ & $0.18$ & $0.19$ \\
  \hline
  \end{tabular}
  \vspace{0.5\baselineskip}
  \caption{Predictive correlations for the simulations shown in Figure 1 in the
    paper; the training population for the genomic prediction model is composed
    by $200$ varieties from $2002$--$2007$ WHEAT data. $\rb$ is the average
    predictive correlation for a given generation, training population size and
    number of causal variants; and $\hF$ is the corresponding average $\F$.
    $\rD$ is the decay curve estimate of $\rb$, and is only available if the
    generation average falls within the span of the decay curve. $\rL$ is the
    corresponding estimate from the linear extrapolation.}
  \label{tab:breeding200}
  \end{center}
\end{table}

\pagebreak

\begin{table}[h]
  \begin{center}
  \begin{tabular}{|c|c|c|c|c|c|}
  \hline
  \textbf{Causal Variants} & \textbf{Generation}
                & $\hF$   & $\rb$  & $\rD$  & $\rL$ \\
  \hline
  $200$  &  $1$ & $0.018$ & $0.58$ & $0.55$ & $0.55$ \\
  \hline
  $200$  &  $2$ & $0.041$ & $0.47$ & $0.51$ & $0.51$ \\
  \hline
  $200$  &  $3$ & $0.066$ & $0.40$ & $-$    & $0.46$ \\
  \hline
  $200$  &  $4$ & $0.088$ & $0.36$ & $-$    & $0.42$ \\
  \hline
  $200$  &  $5$ & $0.111$ & $0.30$ & $-$    & $0.38$ \\
  \hline
  $200$  &  $6$ & $0.127$ & $0.27$ & $-$    & $0.35$ \\
  \hline
  $200$  &  $7$ & $0.141$ & $0.25$ & $-$    & $0.33$ \\
  \hline
  $200$  &  $8$ & $0.151$ & $0.20$ & $-$    & $0.31$ \\
  \hline
  $200$  &  $9$ & $0.158$ & $0.19$ & $-$    & $0.30$ \\
  \hline
  $200$  & $10$ & $0.165$ & $0.15$ & $-$    & $0.28$ \\
  \hline
  $1000$ & $1$  & $0.019$ & $0.62$ & $0.53$ & $0.53$ \\
  \hline
  $1000$ & $2$  & $0.047$ & $0.50$ & $0.48$ & $0.47$ \\
  \hline
  $1000$ & $3$  & $0.077$ & $0.46$ & $-$    & $0.41$ \\
  \hline
  $1000$ & $4$  & $0.106$ & $0.40$ & $-$    & $0.35$ \\
  \hline
  $1000$ & $5$  & $0.126$ & $0.33$ & $-$    & $0.31$ \\
  \hline
  $1000$ & $6$  & $0.139$ & $0.30$ & $-$    & $0.28$ \\
  \hline
  $1000$ & $7$  & $0.150$ & $0.25$ & $-$    & $0.26$ \\
  \hline
  $1000$ & $8$  & $0.157$ & $0.20$ & $-$    & $0.24$ \\
  \hline
  $1000$ & $9$  & $0.164$ & $0.19$ & $-$    & $0.23$ \\
  \hline
  $1000$ & $10$ & $0.168$ & $0.15$ & $-$    & $0.22$ \\
  \hline
  \end{tabular}
  \vspace{0.5\baselineskip}
  \caption{Predictive correlations for the simulations shown in Figure 2 in
    the paper; the training population for the genomic prediction model is
    composed by the $800$ varieties available after the second round of
    selection in the simulation.
    The notation is the same as in Table \ref{tab:breeding200}.}
  \label{tab:breeding800}
  \end{center}
\end{table}

\pagebreak

\subsection{Cross-Population Simulation using the HUMAN data}

\begin{table}[h]
  \begin{center} \scriptsize
  \begin{tabular}{|p{2cm}|p{2cm}|c|c|c|c|c|}
  \hline
  \textbf{Training \newline Population} & \textbf{Target \newline Population} & 
  \textbf{Causal Variants}  & $\hF$ & $\rP$ & $\rD$ & $\rL$ \\
  \hline
  \multirow{5}{*}{Asia}
    & Europe      & $5$     & $0.068$ & $0.68$  & $0.65$ & $0.66$ \\
  \cline{2-7}
    & Middle east & $5$     & $0.076$ & $0.67$  & $0.65$ & $0.65$ \\
  \cline{2-7}
    & America     & $5$     & $0.154$ & $0.69$  & $-$ & $0.62$ \\
  \cline{2-7}
    & Africa      & $5$     & $0.156$ & $0.64$  & $-$ & $0.62$ \\
  \cline{2-7}
    & Oceania     & $5$     & $0.174$ & $0.78$  & $-$ & $0.62$ \\
  \hline
  \multirow{5}{*}{Asia}
    & Europe      & $20$    & $0.068$ & $0.49$  & $0.45$ & $0.45$ \\
  \cline{2-7}
    & Middle east & $20$    & $0.076$ & $0.32$  & $0.39$ & $0.39$ \\
  \cline{2-7}
    & America     & $20$    & $0.154$ & $0.48$  & $-$    & $0.39$ \\
  \cline{2-7}
    & Africa      & $20$    & $0.156$ & $0.59$  & $-$    & $0.45$ \\
  \cline{2-7}
    & Oceania     & $20$    & $0.174$ & $0.43$  & $-$    & $0.37$ \\
  \hline
  \multirow{5}{*}{Asia}
    & Europe      & $100$   & $0.068$ & $0.09$  & $0.17$ & $0.17$ \\
  \cline{2-7}
    & Middle east & $100$   & $0.076$ & $0.12$  & $0.15$ & $0.15$ \\
  \cline{2-7}
    & America     & $100$   & $0.154$ & $0.02$  & $-$    & $0.00$ \\
  \cline{2-7}
    & Africa      & $100$   & $0.156$ & $0.15$  & $-$    & $0.00$ \\
  \cline{2-7}
    & Oceania     & $100$   & $0.174$ & $0.03$  & $-$    & $-0.05$ \\
  \hline
  \multirow{5}{*}{Asia}
    & Europe      & $2000$  & $0.068$ & $0.13$  & $0.08$ & $0.08$ \\
  \cline{2-7}
    & Middle east & $2000$  & $0.076$ & $0.14$  & $0.07$ & $0.07$ \\
  \cline{2-7}
    & America     & $2000$  & $0.154$ & $0.24$  & $-$    & $0.02$ \\
  \cline{2-7}
    & Africa      & $2000$  & $0.156$ & $0.03$  & $-$    & $0.02$ \\
  \cline{2-7}
    & Oceania     & $2000$  & $0.174$ & $0.03$  & $-$    & $0.01$ \\
  \hline
  \multirow{5}{*}{Asia}
    & Europe      & $10000$ & $0.068$ & $0.15$  & $0.10$ & $0.10$ \\
  \cline{2-7}
    & Middle east & $10000$ & $0.076$ & $0.21$  & $0.10$ & $0.10$ \\
  \cline{2-7}
    & America     & $10000$ & $0.154$ & $0.02$  & $-$    & $0.08$ \\
  \cline{2-7}
    & Africa      & $10000$ & $0.156$ & $0.22$  & $-$    & $0.08$ \\
  \cline{2-7}
    & Oceania     & $10000$ & $0.174$ & $-0.18$ & $-$    & $0.08$ \\
  \hline
  \multirow{5}{*}{Asia}
    & Europe      & $50000$ & $0.068$ & $0.28$  & $0.02$ & $0.02$ \\
  \cline{2-7}
    & Middle east & $50000$ & $0.076$ & $0.11$  & $0.01$ & $0.01$ \\
  \cline{2-7}
    & America     & $50000$ & $0.154$ & $0.00$  & $-$    & $-0.07$ \\
  \cline{2-7}
    & Africa      & $50000$ & $0.156$ & $-0.10$ & $-$    & $-0.07$ \\
  \cline{2-7}
    & Oceania     & $50000$ & $0.174$ & $-0.10$ & $-$    & $-0.09$ \\
  \hline
  \end{tabular}
  \vspace{0.5\baselineskip}
  \caption{Predictive correlations for the simulations shown in Figure 3 in the
  paper. $\rP$ is the predictive correlation for the target population from the
  full training population. $\rD$ is the decay curve estimate of $\rP$, and is
  only available if the target population falls within the span of the decay
  curve. $\rL$ is the corresponding estimate from the linear extrapolation.}
  \label{tab:hgdp}
  \end{center}
\end{table}

\pagebreak

\section{Real-World Data Analyses}

\subsection{WHEAT Data}

\begin{figure}[h]
  \begin{center}
  \includegraphics[width=0.75\textwidth]{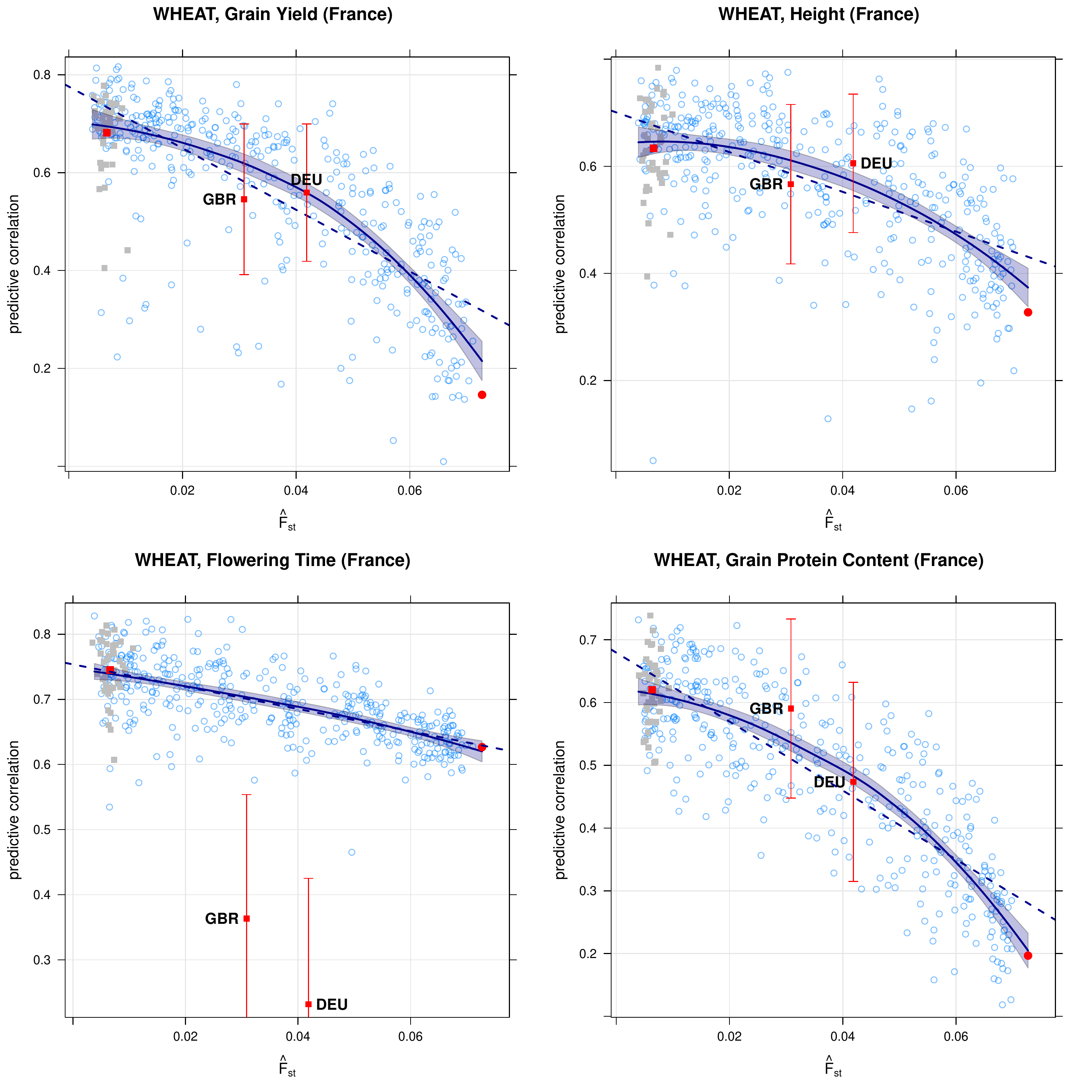}
  \caption{Decay curves for grain yield, height, flowering time and grain
    protein content estimated from the French wheat varieties in the WHEAT data.
    The blue circles are the $\rD^{(m)}$ used to build the curve, and the red
    point is $\rD^{(0)}$. The blue line is the mean decay trend, with a shaded
    $95\%$ confidence interval, and the dashed blue line is the linear
    interpolation provided by the $\rL$. Gray squares are the $\rCV$ computed
    using hold-out cross-validation. The red squares labelled GBR and DEU
    correspond to the $\rP$ for the British and German varieties, and the red
    brackets are the respective $95\%$ confidence intervals.}
  \label{fig:tg-traits}
  \end{center}
\end{figure}

\pagebreak

\subsection{MICE Data}

\begin{figure}[h]
  \begin{center}
  \includegraphics[width=0.8\textwidth]{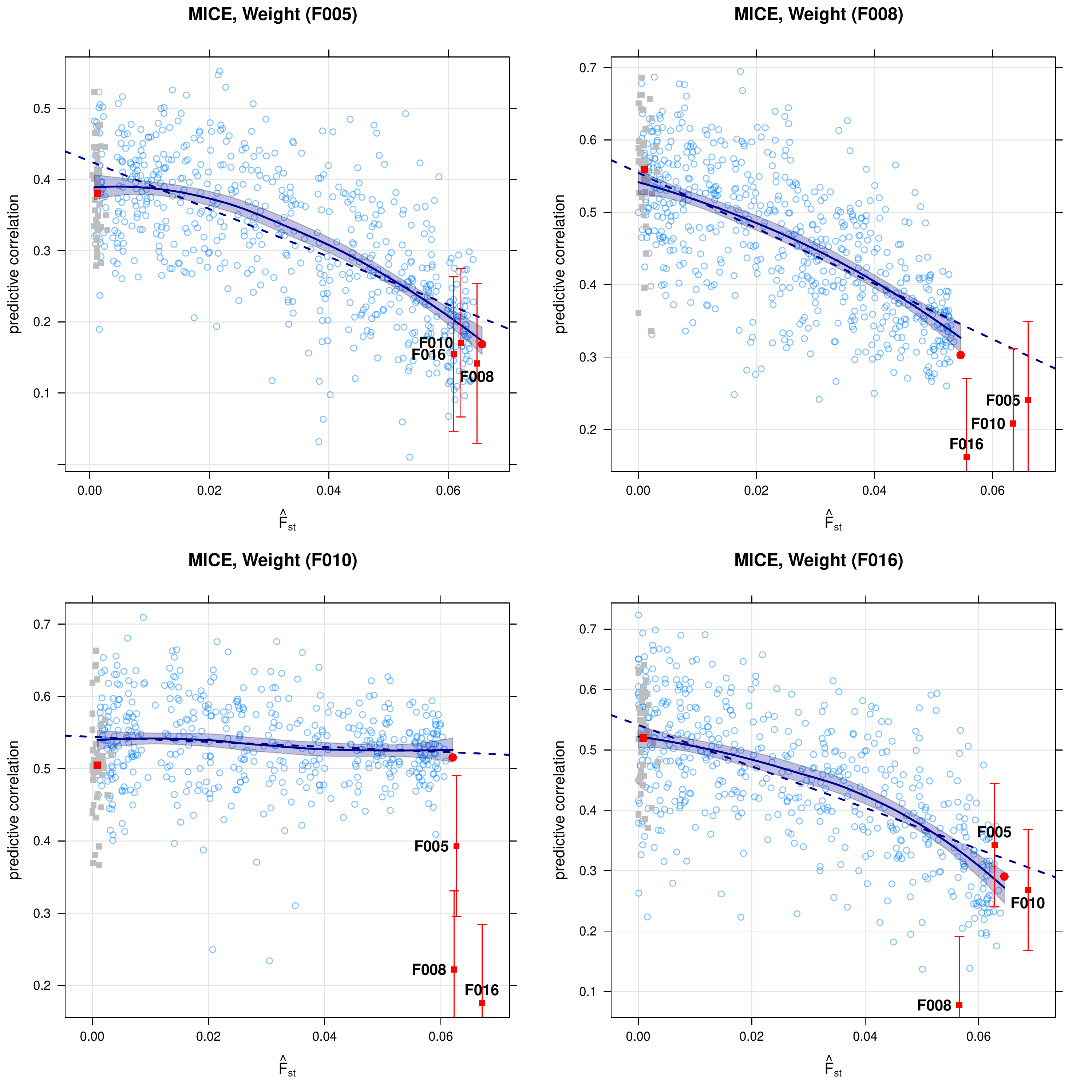}
  \caption{Decay curves for weight estimated from the $4$ largest families in
    the MICE data, labelled F005, F008, F010 and F016. The red squares in each
    panel correspond to the predictive correlations for the populations not
    used for estimating the decay curve; the red brackets are $95\%$ confidence
    intervals. Formatting is the same as in Figure \ref{fig:tg-traits}.}
  \label{fig:mice-weight}
  \end{center}
\end{figure}

\pagebreak

\begin{figure}[h]
  \begin{center}
  \includegraphics[width=0.8\textwidth]{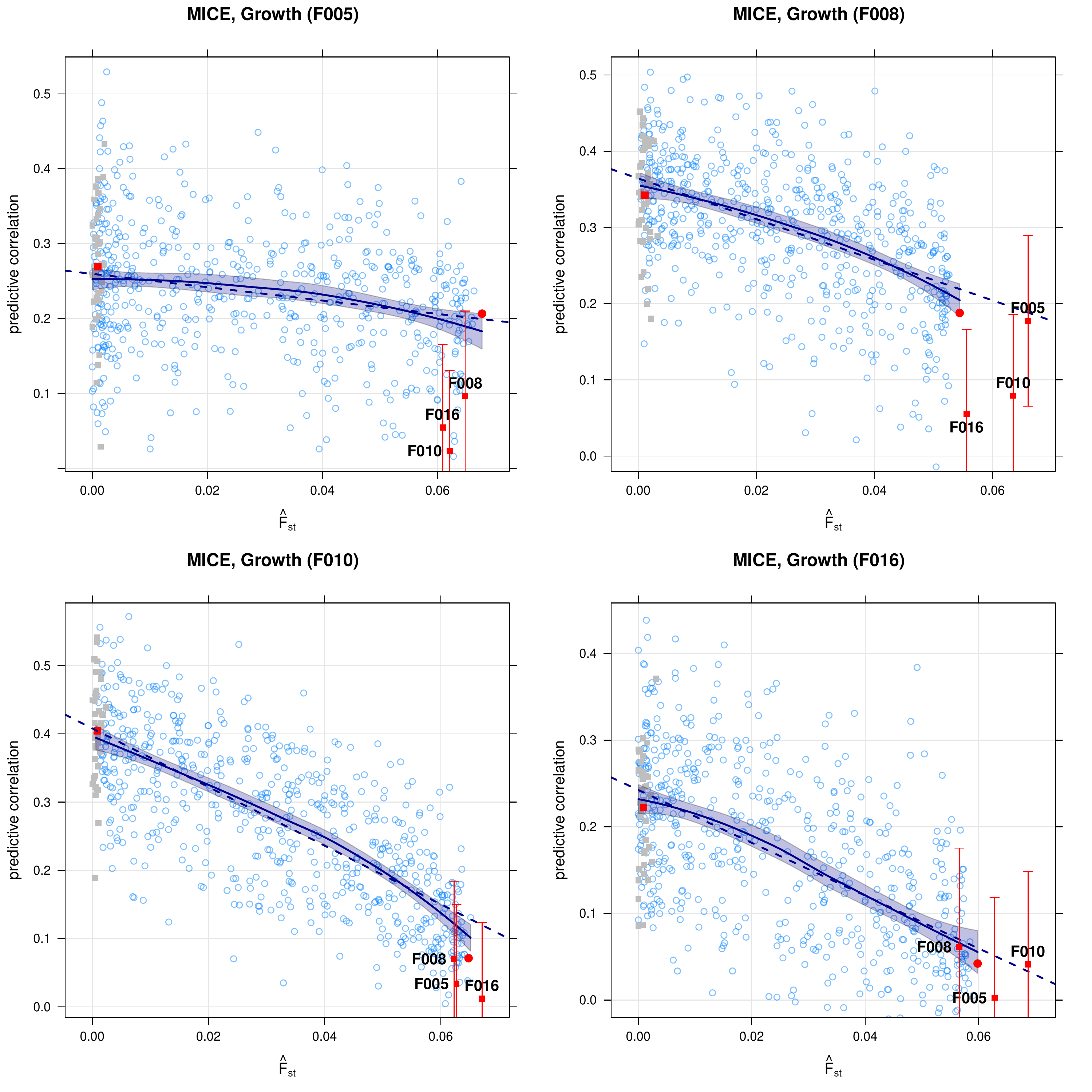}
  \caption{Decay curves for growth rate estimated from the $4$ largest families
    in the MICE data, labelled F005, F008, F010 and F016. The red squares in
    each panel correspond to the predictive correlations for the populations not
    used for estimating the decay curve; the red brackets are $95\%$ confidence
    intervals. Formatting is the same as in Figure \ref{fig:tg-traits}.}
  \label{fig:mice-growth}
  \end{center}
\end{figure}

\pagebreak

\subsection{Cross-Validation and Decay Curve in the WHEAT and MICE data}

\begin{table}[ht!]
  \begin{center}
  \begin{tabular}{|p{3cm}|p{2cm}|c|c|c|}
  \hline
  \textbf{Trait} & \textbf{Training \newline Population} &  
    $\hF$ & $\rCV$ & $\rD$ \\
  \hline
  WHEAT, Yield  & France & $0.006$ & $0.68$ & $0.68$ \\
  \hline
  WHEAT, Height & France & $0.006$ & $0.63$ & $0.64$ \\
  \hline
  \parbox[c][2.5\baselineskip]{3cm}{WHEAT, \newline Flowering time}
                & France & $0.006$ & $0.74$ & $0.74$ \\
  \hline
  \parbox[c][2.5\baselineskip]{3cm}{WHEAT, Grain \newline protein content}
                & France & $0.006$ & $0.62$ & $0.61$ \\
  \hline
  \multirow{4}{*}{MICE, Weight}
                & F005 & $0.001$ & $0.38$ & $0.39$ \\
  \cline{2-5}
                & F008 & $0.001$ & $0.56$ & $0.53$ \\
  \cline{2-5}
                & F010 & $0.001$ & $0.50$ & $0.54$ \\
  \cline{2-5}
                & F016 & $0.001$ & $0.52$ & $0.52$ \\
  \hline
  \multirow{4}{*}{\parbox{3cm}{MICE, \newline Growth rate}}
                & F005 & $0.001$ & $0.27$ & $0.25$ \\
  \cline{2-5}
                & F008 & $0.001$ & $0.34$ & $0.35$ \\
  \cline{2-5}
                & F010 & $0.001$ & $0.40$ & $0.38$ \\
  \cline{2-5}
                & F016 & $0.001$ & $0.22$ & $0.23$ \\
  \hline
  \end{tabular}
  \label{tab:refpoints}
  \vspace{0.5\baselineskip}
  \caption{Predictive correlations from the decay curves and from
    cross-validation for the analyses shown in Figures \ref{fig:tg-traits},
    \ref{fig:mice-weight} and \ref{fig:mice-growth}. $\hF$ and $\rCV$ are the
    mean genetic distance and mean predictive correlation from the $40$ runs of
    hold-out cross-validation; $\rD$ is the predictive correlation estimated
    by the decay curve at genetic distance $\hF$.}
  \end{center}
\end{table}

\pagebreak

\section{Kinship and $\F$}

\begin{table}[h]
  \begin{center}
  \begin{tabular}{|r|r|c|c|c|}
  \hline
  Data      & Subset & $m$s & $\COR(\hF^{(m)}, \bk^{(m)})$ & $log_{10}(p)$ \\
  \hline
  WHEAT & France & 401 & $-0.9894$ & $-672.10$ \\
  MICE  & F005   & 601 & $-0.9982$ & $-1467.58$ \\
  MICE  & F008   & 601 & $-0.9982$ & $-1467.58$\\
  MICE  & F010   & 601 & $-0.9906$ & $-1038.57$ \\
  MICE  & F016   & 601 & $-0.9948$ & $-1192.05$ \\
  HUMAN & Asia   & 601 & $-0.9998$ & $-2038.97$ \\
  \hline
  \end{tabular}
  \vspace{0.5\baselineskip}
  \caption{Correlation between $\hF^{(m)}$ and $\bk^{(m)}$ in the data sets and
    training populations used in the paper. The p-values are computed using the
    exact t-test for the correlation coefficient \cite{hotelling} and adjusted
    for multiplicity via FDR \cite{fdr}.}
  \end{center}
\end{table}

\begin{figure}[h]
  \begin{center}
  \includegraphics[width=0.55\textwidth]{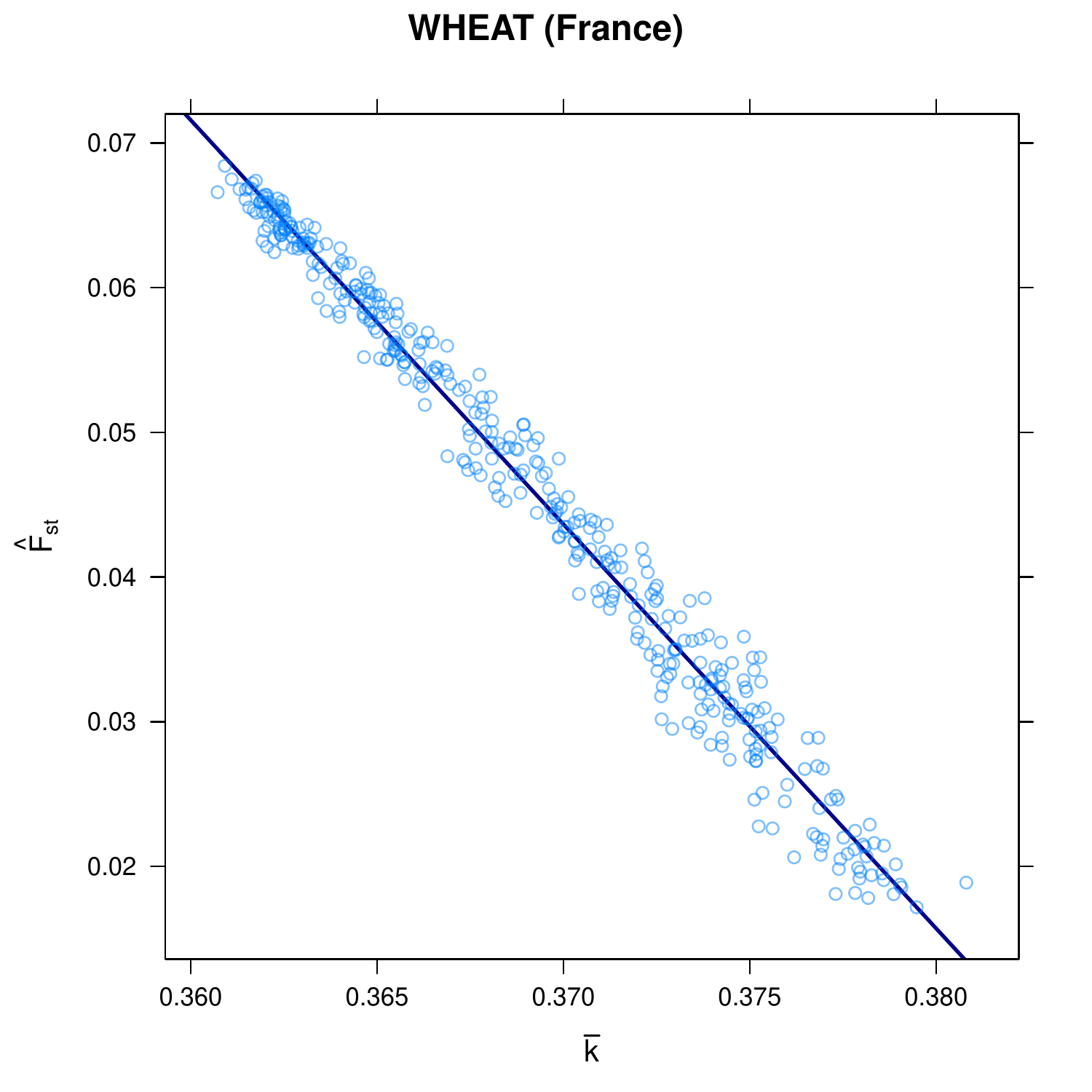}
  \caption{$(\F^{(m)}, \bk^{(m)})$ pairs generated from the French wheat
    varieties in the WHEAT data.}
  \label{fig:tg-kfst}
  \end{center}
\end{figure}

\pagebreak

\begin{figure}[h]
  \begin{center}
  \includegraphics[width=0.8\textwidth]{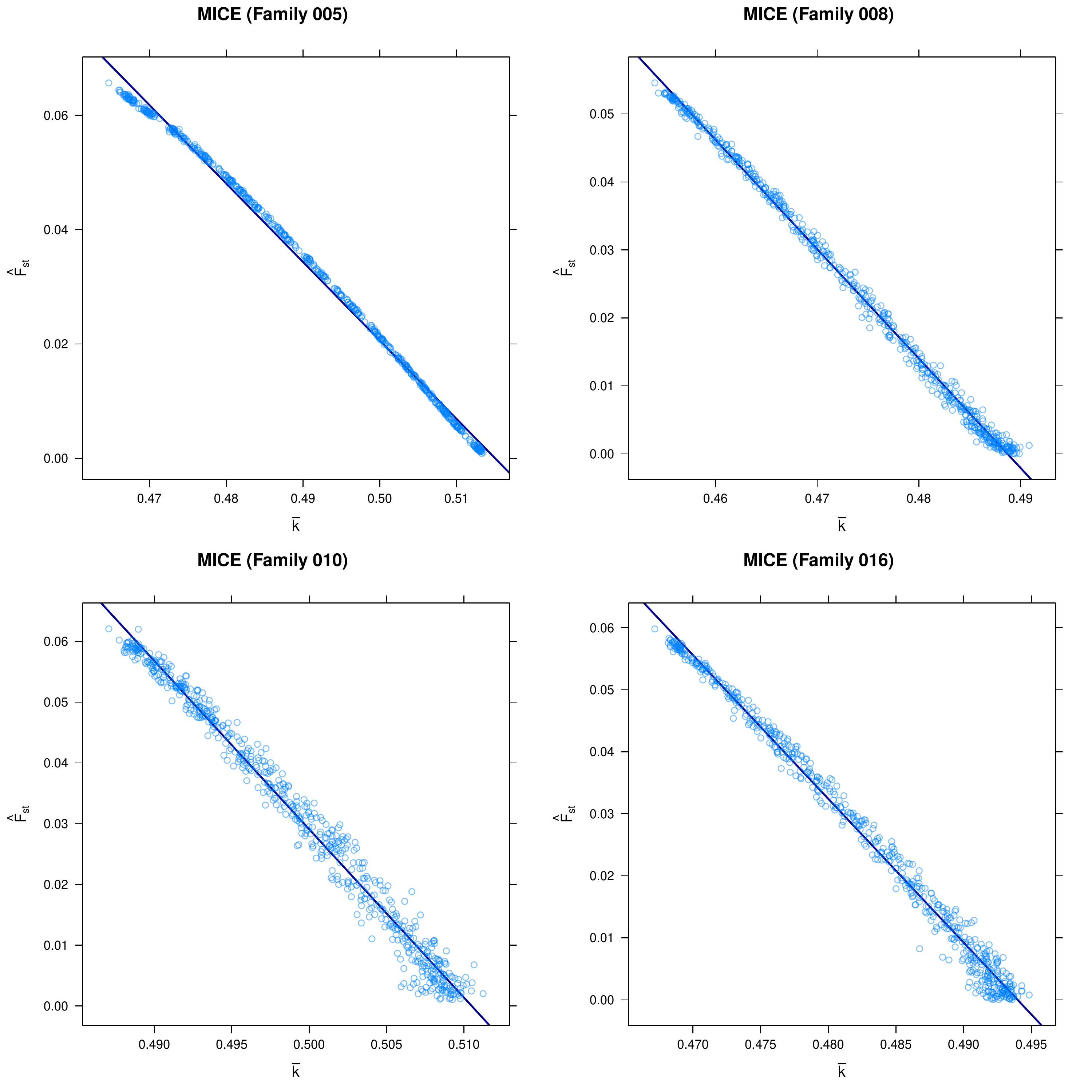}
  \caption{$(\F^{(m)}, \bk^{(m)})$ pairs generated from the $4$ largest
    families in the MICE data, labelled F005, F008, F010 and F016.}
  \label{fig:mice-kfst}
  \end{center}
\end{figure}

\pagebreak

\begin{figure}[h]
  \begin{center}
  \includegraphics[width=0.8\textwidth]{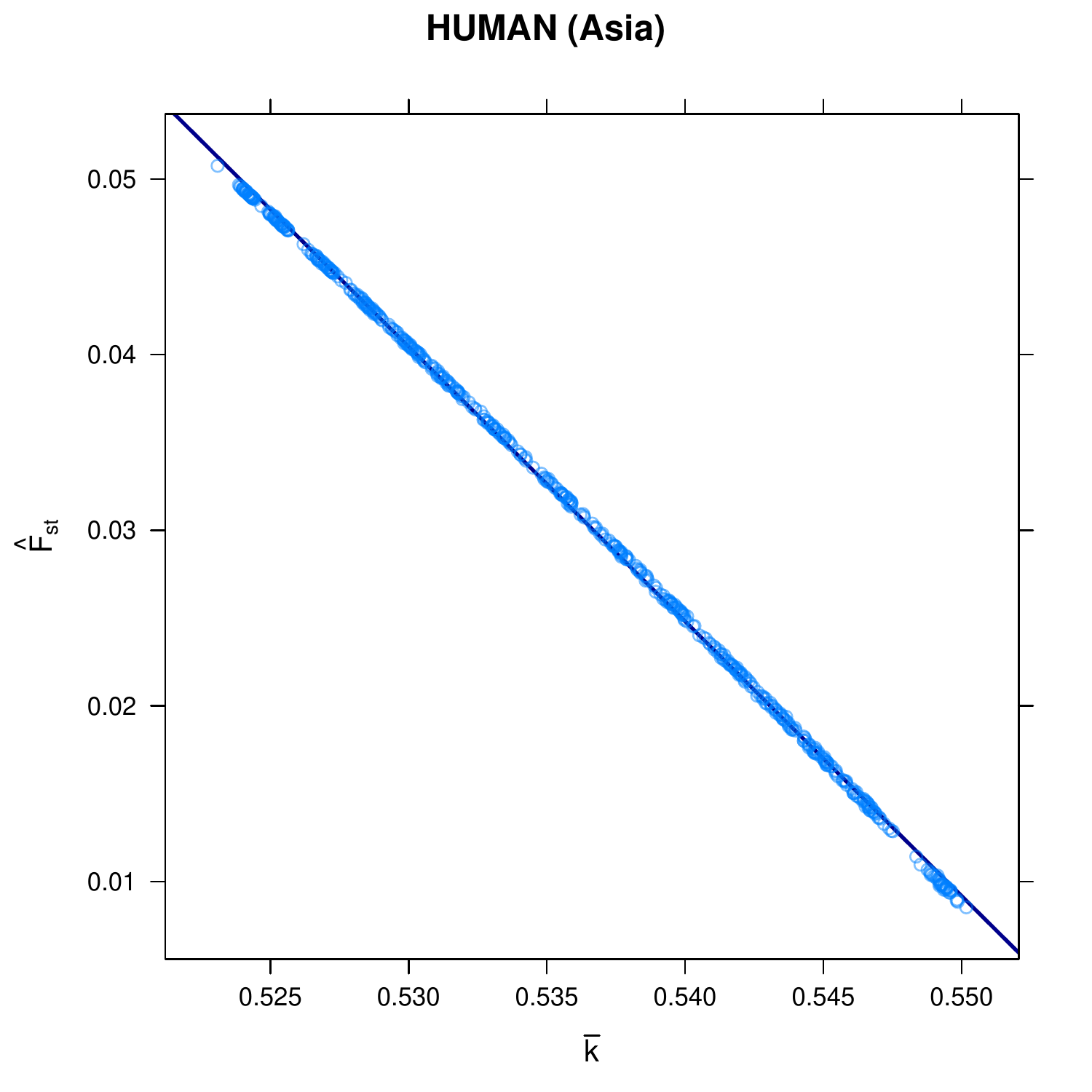}
  \caption{$(\F^{(m)}, \bk^{(m)})$ pairs generated from the Asian individuals
    in the HUMAN data.}
  \label{fig:hgdp-kfst}
  \end{center}
\end{figure}

\clearpage
\pagebreak

\section{Relationship between Squared Predictive Correlation and $\F^2$}

\cite{pszczola} used a simulated dairy cattle population, created simulating
both phenotypes and genotypes, suggested that squared predictive correlation
has a stronger linear relationship with squared mean kinship than predictive
correlation does with mean kinship. Predictive correlation was computed
using GBLUP as a genomic prediction model.

In the context of this paper, this is equivalent to testing whether the 
$\left(\rD^{(m)}\right)^2$ have a stronger linear relationship with the
$\left(\hF^{(m)}\right)^2$ than the $\rD^{(m)}$ do with the $\hF^{(m)}$; we
have shown that $\F^{(m)}$ and $\bk^{(m)}$ are almost perfectly linearly
correlated so they can be used interchangeably for this purpose. We regress the
$\rD^{(m)}$ on the $\hF^{(m)}$ and measure the $R^2$ coefficient of the
resulting linear model, denoted as $\mathrm{R^2}_{\mathrm{LINEAR}}$. Similarly,
we regress the $\left(\rD^{(m)}\right)^2$ on the $\left(\hF^{(m)}\right)^2$
and measure $\mathrm{R^2}_{\mathrm{QUADRATIC}}$. Both are reported in Tables
\ref{tab:r2real} and \ref{tab:r2sim} for all the analyses with real and
simulated phenotypes.

To test whether there is a significant difference between
$\mathrm{R^2}_{\mathrm{LINEAR}}$ and $\mathrm{R^2}_{\mathrm{QUADRATIC}}$ we
perform a permutation two-sample $t$-test as described in \cite{pesarin}, using
$10000$ permutations. The resulting p-value is $0.784$, hence we conclude that
the difference between the relationship we consider in this paper and that 
suggested in \cite{pszczola} is not significant.

\clearpage
\pagebreak

\begin{table}[h!]
  \begin{center}
  \begin{tabular}{|c|c|c|c|c|}
    \hline
    \textbf{Data} & \textbf{Trait} & \textbf{Training Population} & 
      $\mathbf{R^2}_{\mathrm{LINEAR}}$ & $\mathbf{R^2}_{\mathrm{QUADRATIC}}$ \\
    \hline
    \multirow{4}{*}{WHEAT} & Yield                   & France & $0.575$ & $0.634$ \\
    \cline{2-5}
                           & Height                  & France & $0.371$ & $0.424$ \\
    \cline{2-5}
                           & Flowering Time          & France & $0.412$ & $0.410$ \\
    \cline{2-5}
                           & Grain protein content   & France & $0.681$ & $0.681$ \\
    \cline{2-5}
    \hline
    \multirow{8}{*}{MICE}  & \multirow{4}{*}{Weight} & F005   & $0.056$ & $0.064$ \\
    \cline{3-5}
                           &                         & F008   & $0.246$ & $0.236$ \\
    \cline{3-5}
                           &                         & F010   & $0.537$ & $0.463$ \\
    \cline{3-5}
                           &                         & F016   & $0.311$ & $0.242$ \\
    \cline{2-5}
                           & \multirow{4}{*}{Growth} & F005   & $0.446$ & $0.437$ \\
    \cline{3-5}
                           &                         & F008   & $0.426$ & $0.404$ \\
    \cline{3-5}
                           &                         & F010   & $0.013$ & $0.019$ \\
    \cline{3-5}
                           &                         & F016   & $0.384$ & $0.372$ \\
    \hline
  \end{tabular}
  \vspace{0.5\baselineskip}
  \caption{$\mathrm{R^2}_{\mathrm{LINEAR}}$ and $\mathrm{R^2}_{\mathrm{QUADRATIC}}$
    for the data analyses on real phenotypes.}
  \label{tab:r2real}
  \end{center}
\end{table}

\begin{table}[h!]
  \begin{center}
  \begin{tabular}{|c|c|c|c|c|}
    \hline
    \textbf{Simulation} & \textbf{Sample Size} & \textbf{Causal Variants} &
      $\mathbf{R^2}_{\mathrm{LINEAR}}$ & $\mathbf{R^2}_{\mathrm{QUADRATIC}}$ \\
    \hline
    \multirow{6}{*}{Genomic selection} & $200$ & $10$   & $0.387$ & $0.358$ \\
    \cline{2-5}
                                       & $200$ & $50$   & $0.307$ & $0.307$ \\
    \cline{2-5}
                                       & $200$ & $200$  & $0.122$ & $0.112$ \\
    \cline{2-5}
                                       & $200$ & $1000$ & $0.263$ & $0.261$ \\
    \cline{2-5}
                                       & $800$ & $800$  & $0.284$ & $0.293$ \\
    \cline{2-5}
                                       & $800$ & $1000$ & $0.351$ & $0.352$ \\
    \hline
    \multirow{6}{*}{Cross-population}  & $435$ & $5$     & $0.123$ & $0.093$ \\
    \cline{2-5}
                                       & $435$ & $20$    & $0.175$ & $0.167$ \\
    \cline{2-5}
                                       & $435$ & $100$   & $0.565$ & $0.496$ \\
    \cline{2-5}
                                       & $435$ & $2000$  & $0.131$ & $0.116$ \\
    \cline{2-5}
                                       & $435$ & $10000$ & $0.023$ & $0.035$ \\
    \cline{2-5}
                                       & $435$ & $50000$ & $0.256$ & $0.118$ \\
    \hline
  \end{tabular}
  \vspace{0.5\baselineskip}
  \caption{$\mathrm{R^2}_{\mathrm{LINEAR}}$ and $\mathrm{R^2}_{\mathrm{QUADRATIC}}$
    for the data used in the simulation studies.}
  \label{tab:r2sim}
  \end{center}
\end{table}

\end{document}